\documentclass[useAMS,usegraphicx,usenatbib,a4paper]{mn2e}

\def\lesssim{\mathrel{\hbox{\rlap{\hbox{\lower4pt\hbox{$\sim$}}}\hbox{$<$}}}}
\let\la=\lesssim
\def\gtrsim{\mathrel{\hbox{\rlap{\hbox{\lower4pt\hbox{$\sim$}}}\hbox{$>$}}}}
\let\ga=\gtrsim

\def\dm15{$\Delta m_{15}(B)$}

\def\degr{$^\circ$}
\def\kms{\mbox{${\rm km}\:{\rm s}^{-1}\:$}}

\def\OIa{[O\,{\sc i}]}

\def\MgIa{Mg\,{\sc i}]}

\title[Nebular SN Ib/c line profiles]{Nebular emission-line profiles of Type Ib/c Supernovae -- probing the ejecta asphericity.} 
\author[Taubenberger et al.]{S. Taubenberger$^{1}$
\thanks{E-mail: tauben@mpa-garching.mpg.de},
S. Valenti$^{2,3}$, S. Benetti$^{4}$, E. Cappellaro$^{4}$, M. Della Valle$^{5}$, 
\newauthor N. Elias-Rosa$^{1,6}$, S. Hachinger$^{1}$, W. Hillebrandt$^{1}$, K. Maeda$^{1,7}$, P. A. Mazzali$^{1,4}$, 
\newauthor A. Pastorello$^{3}$, F. Patat$^{2}$, S. A. Sim$^{1}$ and M. Turatto$^{8}$\\
$^{1}$Max-Planck-Institut f\"{u}r Astrophysik, Karl-Schwarzschild-Str. 1, 85741 Garching bei M\"{u}nchen, Germany\\
$^{2}$European Southern Observatory (ESO), Karl-Schwarzschild-Str. 2, 85748 Garching bei M\"{u}nchen, Germany\\
$^{3}$Astrophysics Research Centre, School of Mathematics and Physics, Queen's University Belfast, Belfast BT7 1NN, UK\\
$^{4}$INAF Osservatorio Astronomico di Padova, Vicolo dell'Osservatorio 5, 35122 Padova, Italy\\
$^{5}$INAF Osservatorio Astronomico di Capodimonte, Via Moiariello 16, 80131 Napoli, Italy\\
$^{6}$Spitzer Science Center, California Institute of Technology, 1200 E. California Blvd., Pasadena, CA 91125, USA\\
$^{7}$Institute for the Physics and Mathematics of the Universe, Univ. of Tokyo, Kashiwano-ha 5-1-5, Chiba-ken 277-8582, Japan\\
$^{8}$INAF Osservatorio Astrofisico di Catania, Via S.Sofia 78, 95123 Catania, Italy}
\begin{document}

\date{Accepted 2009 April 29. Received 2009 April 21; in original form 2008 October 1}

\pagerange{\pageref{firstpage}--\pageref{lastpage}} \pubyear{2008}

\maketitle

\label{firstpage}

\begin{abstract}
In order to assess qualitatively the ejecta geometry of 
stripped-envelope core-collapse supernovae, we investigate 98 
late-time spectra of 39 objects, many of them previously unpublished. 
We perform a Gauss-fitting of the \OIa\ $\lambda\lambda6300,6364$ 
feature in all spectra, with the position, full width at half maximum 
(FWHM) and intensity of the $\lambda6300$ Gaussian as free parameters, 
and the $\lambda6364$ Gaussian added appropriately to account for the 
doublet nature of the \OIa\ feature. On the basis of the best-fit 
parameters, the objects are organised into morphological classes, and 
we conclude that at least half of all Type Ib/c supernovae must be 
aspherical. Bipolar jet-models do not seem to be universally applicable, 
as we find too few symmetric double-peaked \OIa\ profiles. In some 
objects the \OIa\ line exhibits a variety of shifted secondary peaks 
or shoulders, interpreted as blobs of matter ejected at high velocity 
and possibly accompanied by neutron-star kicks to assure momentum 
conservation. At phases earlier than $\sim$\,$200$\,d, a systematic 
blueshift of the \OIa\ $\lambda\lambda6300,6364$ line centroids can be 
discerned. Residual opacity provides the most convincing explanation 
of this phenomenon, photons emitted on the rear side of the SN being 
scattered or absorbed on their way through the ejecta. Once modified 
to account for the doublet nature of the oxygen feature, the profile 
of \MgIa\ $\lambda4571$ at sufficiently late phases generally 
resembles that of \OIa\ $\lambda\lambda6300,6364$, suggesting 
negligible contamination from other lines and confirming that O and 
Mg are similarly distributed within the ejecta.
\end{abstract}

\begin{keywords}
supernovae: general -- techniques: spectroscopic -- line: profile
\end{keywords}

\section{Introduction}
\label{Introduction}

The geometry of stripped-envelope core-collapse supernova (CC-SN) ejecta has 
been scrutinised for about ten years, since the association of SN~1998bw with 
the nearby $\gamma$-ray burst GRB 980425 \citep{Galama98}. 
Together with subsequent examples of SN-GRB associations (see e.g. 
\citealt{Ferrero06} for an overview) this suggested that at least some Type Ib/c 
supernovae (SNe~Ib/c) may be powered by the same engine as long-duration GRBs, 
and thus that their ejecta may show large-scale asphericity along an axis defined 
by the GRB-jet. Indeed, the nebular spectra of SN~1998bw exhibited properties which 
could not be explained with spherical symmetry \citep{Mazzali01,Maeda02}. Instead, 
a model with high-velocity Fe-rich material ejected along the jet axis, and 
lower-velocity O in a torus perpendicular to this axis, was proposed. From this 
geometry a strong viewing-angle dependence of nebular line profiles was obtained 
\citep{Maeda02}. Of particular note to this study, this model suggests that 
double-peaked O lines should be observed if viewed from a direction perpendicular 
to the jet.

Such a profile was first observed in SN~2003jd \citep{Mazzali05}, an important 
step towards a coherent picture. However, whether large-scale asphericity is 
found only in SNe Ic connected with a GRB (probably only a few percent of all 
SNe Ic; \citealt{Podsiadlowski04,Guetta07}), and how large the degree of 
asphericity in `normal' SNe~Ib/c actually is, are debated. \citet{Maeda08} 
recently studied a sample of nebular spectra of $18$ stripped-envelope CC-SNe and 
found a large fraction of double-peaked [O\,\textsc{i}] $\lambda\lambda6300,6364$ 
line profiles, consistent with about half of all SNe~Ib/c being strongly, or all 
of them moderately, aspherical. Similarly, \citet{Modjaz08} analysed late-time 
spectra of $8$ stripped CC-SNe, concluding that asphericity is ubiquitous in all 
these events, not only the hyperenergetic ones. It should be noted that the Maeda 
et al. and Modjaz et al. works focus on deviations from sphericity on \textit{global} 
scales, as opposed to small-scale clumpiness of the ejecta that results in 
fine-structured emission lines and requires fairly high-resolution nebular spectra 
to be studied (see e.g. \citealt{Filippenko89,Spyromilio94,Matheson00b}).

In this work we conduct a study similar to that of \citet{Maeda08} and 
\citet{Modjaz08}, but based on a larger sample of SNe, considering virtually all 
nebular SN~Ib/c spectra we could access from the literature, complemented by 
$26$ previously unpublished spectra from the Asiago Supernova Archive and recent 
observations carried out at the ESO Very Large Telescope (VLT). The work is 
organised as follows: 
in Section~\ref{sample of spectra} we present the entire SN sample, discuss the 
selection criteria for spectra to be included, and discuss in more detail those 
spectra which were previously unpublished. Section~\ref{Fitting the oxygen line} 
concentrates on the link between the ejecta geometry and observed line profiles, 
motivates the choice to focus on [O\,\textsc{i}] $\lambda\lambda6300,6364$, 
addresses the complications arising from its doublet nature, and introduces the 
fitting procedure employed to gain qualitative insight into the ejecta morphology. 
The results of this fitting are analysed in Sections~\ref{blueshift} and 
\ref{statistical analysis}, trying to find an explanation for the mean 
blueshifts of the line's centroids at phases $\la 200$\,d, and dividing the 
objects into different classes on the basis of the best-fit parameters. 
Individual objects with interesting line profiles are discussed more deeply in 
Section~\ref{Discussion of individual objects}, while Section~\ref{Mg profile} 
extends the analysis to the profile of Mg\,\textsc{i}] $\lambda4571$ and its 
comparison to that of [O\,\textsc{i}]. Finally, a brief summary of the main 
results is given in Section~\ref{Conclusions}.

\section{The sample of SN~I\lowercase{b/c} spectra}
\label{sample of spectra}

Our goal is to compare a large set of late-time spectra of stripped-envelope 
CC-SNe, concentrating on what can be learned about the ejecta geometry by 
studying the profiles of nebular emission lines, in particular [O\,\textsc{i}] 
$\lambda\lambda6300,6364$ (the motivation to focus on this line is given in 
Section~\ref{Fitting the oxygen line}). 

Given the statistical approach of this study, we use a simple fitting procedure 
(for details see Section~\ref{Fitting the oxygen line}). Compared to full spectral 
modelling this method has the advantage of being fast, capable of dealing with 
complex profiles, and independent of an accurate flux calibration of the spectra, 
thus allowing it to be applied to a large number of spectra.

Nebular emission features in SNe~Ib/c typically start to emerge about two months 
past maximum light, but at that epoch the SN flux is still dominated by photospheric 
emission. For this reason we included in our sample only spectra which were taken 
more than $\sim$\,$90$\,d after maximum light, which, assuming typical rise times, 
corresponds to $100$ or $110$\,d after explosion. At those phases there still is an 
underlying photospheric continuum, but this should not affect severely the profiles 
of forbidden emission lines, so that they can be used to trace the geometry of the 
ejecta (but see Section~\ref{blueshift} for the consequence of residual optical 
depth). \citet{Maeda08} employed a more stringent criterion, restricting their sample 
to spectra with epochs $\geq$\,$200$\,d to avoid any possible deformation of lines 
by optical-depth effects. This criterion would reduce our sample from $98$ 
to $53$ spectra (including $16$ SNe not analysed by Maeda et al.), and deprive us 
of the possibility to investigate when the ejecta become fully transparent, which 
is addressed in Section~\ref{blueshift}.

Applying a phase cut required a fairly precise estimate of the epoch of explosion or 
maximum light. Whenever a complete light curve was not available, this information 
was reconstructed from discovery and classification remarks reported in 
IAU-circulars. In exception to this rule, spectra of three SNe were included for 
which no early observations exist. However, the spectrum of SN~1995bb 
\citep{Matheson01} is decidedly nebular, as are the later two out of three spectra 
of SN~1990aj. The first one in this series appears rather peculiar and may still 
show some photospheric features, but was included for completeness. Finally, a 
spectrum of SN~2005N was dated to $\sim$\,$90$\,d past maximum light by 
cross-correlation with a set of comparison spectra \citep{Harutyunyan07}. Besides 
the constraints on the phase, also spectra with insufficient signal-to-noise ratio 
(S/N) in the wavelength region of interest were rejected.

\subsection{The full sample}
\label{The full sample}

\begin{table*}
\caption[List of SNe~Ib/c studied in this paper.]{List of SNe~Ib/c included in the sample -- $1^{st}$ part.} 
\label{spectra}
\center
\begin{scriptsize}
\begin{tabular}{llllccccl}
\hline \\[-1.7ex]
SN     & Type           & Host galaxy  & Morphology$^a$ & $v_\mathrm{rec}^b$ & Redshift & Date           & Epoch$^c$  & Reference \\[0.3ex]
\hline  \\[-1.7ex]
1983N  & Ib             & NGC 5236     & SBc            &  $554\pm119$   & 0.0018(04)   & 1984/03/01     & $226\pm 5$ & \citealt{Gaskell86}\\
1985F  & Ib/c           & NGC 4618     & SBm            &  $544\pm 59$   & 0.0018(02)   & 1985/03/19     & $280\pm 4$ & \citealt{Filippenko86}\\
1987M  & Ic             & NGC 2715     & SABc           & $1216\pm132$   & 0.0041(04)   & 1988/02/09     & $141\pm 7$ & \citealt{Filippenko90}\\
       &                &              &                &                &              & 1988/02/25     & $157\pm 7$ & \citealt{Filippenko90}\\
1988L  & Ic             & NGC 5480     & Sc             & $1963\pm100$   & 0.0065(03)   & 1988/07/17     &  $90\pm12$ & \citealt{Matheson01}\\
       &                &              &                &                &              & 1988/09/15     & $149\pm12$ & \citealt{Matheson01}\\
1990B  & Ic             & NGC 4568     & Sbc(M)         & $2255\pm153$   & 0.0075(05)   & 1990/04/19     &  $91\pm 2$ & \citealt{Clocchiatti01}\\
       &                &              &                &                &              & 1990/04/30     & $102\pm 2$ & \citealt{Matheson01}\\
1990I  & Ib             & NGC 4650A    & S0/a(M)        & $2880\pm 99^d$ & 0.0096(03)   & 1990/07/26     &  $90\pm 2$ & \citealt{Elmhamdi04}\\
       &                &              &                &                &              & 1990/12/21     & $237\pm 2$ & \citealt{Elmhamdi04}\\
       &                &              &                &                &              & 1991/02/20     & $298\pm 2$ & \citealt{Elmhamdi04}\\
1990U  & Ic             & NGC 7479     & SBbc           & $2525\pm162$   & 0.0084(05)   & 1990/10/20     & $100\pm12$ & \citealt{Matheson01}\\
       &                &              &                &                &              & 1990/10/24     & $104\pm12$ & \citealt{Gomez94}\\
       &                &              &                &                &              & 1990/11/23     & $134\pm12$ & Asiago archive\\
       &                &              &                &                &              & 1990/11/28     & $139\pm12$ & \citealt{Matheson01}\\
       &                &              &                &                &              & 1990/12/12     & $153\pm12$ & \citealt{Matheson01}\\
       &                &              &                &                &              & 1990/12/20     & $161\pm12$ & Asiago archive\\
       &                &              &                &                &              & 1991/01/06     & $178\pm12$ & \citealt{Matheson01}\\
       &                &              &                &                &              & 1991/01/12     & $184\pm12$ & \citealt{Gomez94}\\
1990W  & Ib/c           & NGC 6221     & SBc            & $1481\pm126$   & 0.0049(04)   & 1991/02/21     & $183\pm 3$ & Asiago archive\\
       &                &              &                &                &              & 1991/04/21     & $242\pm 3$ & Asiago archive\\
1990aa & Ic             & MCG+05-03-016& Sb             & $5032\pm108$   & 0.0168(04)   & 1991/01/12     & $130\pm 7$ & \citealt{Gomez02}\\
       &                &              &                &                &              & 1991/01/23     & $141\pm 7$ & \citealt{Matheson01}\\
1990aj & Ib/c$_{pec}$   & NGC 1640     & SBb(R)         & $1604\pm 64^d$ & 0.0053(02)   & 1991/01/29     & $140\pm50$ & Asiago archive\\
       &                &              &                &                &              & 1991/02/22     & $164\pm50$ & Asiago archive\\
       &                &              &                &                &              & 1991/03/10     & $180\pm50$ & \citealt{Matheson01}\\
1991A  & Ic             & IC 2973      & SBcd           & $3232\pm 84$   & 0.0107(03)   & 1991/03/22     &  $99\pm10$ & \citealt{Gomez94}\\
       &                &              &                &                &              & 1991/04/07     & $115\pm10$ & \citealt{Matheson01}\\
       &                &              &                &                &              & 1991/04/16     & $124\pm10$ & \citealt{Gomez94}\\
       &                &              &                &                &              & 1991/06/08     & $177\pm10$ & \citealt{Gomez94}\\
1991L  & Ib/c           & MCG+07-34-134& Sc(M)          & $9186\pm200^d$ & 0.0306(07)   & 1991/06/08     & $121\pm20$ & \citealt{Gomez02}\\
1991N  & Ic             & NGC 3310     & SABb(R)        & $1071\pm 80$   & 0.0036(03)   & 1991/12/14     & $274\pm15$ & \citealt{Matheson01}\\
       &                &              &                &                &              & 1992/01/09     & $300\pm15$ & \citealt{Matheson01}\\
1993J  & IIb            & NGC 3031     & Sab            & $-140\pm192^d$ &-0.0001(06)   & 1993/10/19     & $205\pm 3$ & \citealt{Barbon95}\\
       &                &              &                &                &              & 1993/11/19     & $236\pm 3$ & \citealt{Barbon95}\\
       &                &              &                &                &              & 1993/12/08     & $255\pm 3$ & \citealt{Barbon95}\\
       &                &              &                &                &              & 1994/01/17     & $295\pm 3$ & \citealt{Barbon95}\\
       &                &              &                &                &              & 1994/01/21     & $299\pm 3$ & \citealt{Barbon95}\\
       &                &              &                &                &              & 1994/01/22     & $300\pm 3$ & \citealt{Barbon95}\\
       &                &              &                &                &              & 1994/03/25     & $362\pm 3$ & \citealt{Barbon95}\\
       &                &              &                &                &              & 1994/03/30     & $367\pm 3$ & \citealt{Barbon95}\\
1994I  & Ic             & NGC 5194     & Sbc(M)         &  $493\pm 70$   & 0.0016(02)   & 1994/07/14     &  $97\pm 1$ & \citealt{Filippenko95}\\
       &                &              &                &                &              & 1994/08/04     & $118\pm 1$ & \citealt{Filippenko95}\\
       &                &              &                &                &              & 1994/09/02     & $147\pm 1$ & \citealt{Filippenko95}\\
1995bb & Ib/c           & anonymous    & S/Irr          & $1626\pm250$   & 0.0054(08)   & 1995/12/17     & nebular    & \citealt{Matheson01}\\
1996D  & Ic             & NGC 1614     & SBc(M)         & $4531\pm167$   & 0.0151(06)   & 1996/09/10     & $214\pm10$ & Asiago archive\\
1996N  & Ib             & NGC 1398     & SBab(R)        & $1396\pm214$   & 0.0047(07)   & 1996/10/19     & $224\pm 7$ & \citealt{Sollerman98}\\
       &                &              &                &                &              & 1996/12/16     & $282\pm 7$ & \citealt{Sollerman98}\\
       &                &              &                &                &              & 1997/01/13     & $310\pm 7$ & \citealt{Sollerman98}\\
       &                &              &                &                &              & 1997/02/12     & $340\pm 7$ & \citealt{Sollerman98}\\
1996aq & Ib             & NGC 5584     & SABc           & $1675\pm 83$   & 0.0056(03)   & 1997/02/11     & $176\pm 4$ & Asiago archive\\
       &                &              &                &                &              & 1997/04/02     & $226\pm 4$ & Asiago archive\\
       &                &              &                &                &              & 1997/05/14     & $268\pm 4$ & Asiago archive\\
1997B  & Ic             & IC 438       & SABc(R)        & $2919\pm144$   & 0.0097(05)   & 1997/09/23     & $262\pm 5$ & Asiago archive\\
       &                &              &                &                &              & 1997/10/11     & $280\pm 5$ & Asiago archive\\
       &                &              &                &                &              & 1998/02/02     & $394\pm 5$ & Asiago archive\\
1997X  & Ic             & NGC 4691     & SB0/a          & $1072\pm 47$   & 0.0036(02)   & 1997/05/10     & $103\pm 5$ & \citealt{Gomez02}\\
       &                &              &                &                &              & 1997/05/13     & $106\pm 5$ & Asiago archive\\
1997dq & Ic             & NGC 3810     & Sc             &  $993\pm114^d$ & 0.0033(04)   & 1998/05/30     & $217\pm10$ & Asiago archive\\
       &                &              &                &                &              & 1998/06/18     & $236\pm10$ & \citealt{Matheson01}\\
1997ef & BL-Ic          & UGC 4107     & Sc             & $3452\pm 63$   & 0.0115(02)   & 1998/09/21     & $287\pm 3$ & \citealt{Matheson01}\\
1998bw & BL-Ic          & ESO184-G82   & SBbc           & $2445\pm135$   & 0.0082(05)   & 1998/09/12     & $126\pm 1$ & \citealt{Patat01}\\
       &                &              &                &                &              & 1998/11/26     & $201\pm 1$ & \citealt{Patat01}\\
       &                &              &                &                &              & 1999/04/12     & $337\pm 1$ & \citealt{Patat01}\\
       &                &              &                &                &              & 1999/05/21     & $376\pm 1$ & \citealt{Patat01}\\
1999cn & Ic             & MCG+02-38-043& Sab            & $6502\pm235$   & 0.0217(08)   & 2000/04/08     & $297\pm 5$ & Asiago archive\\
1999dn & Ib             & NGC 7714     & SBb(M)         & $2744\pm 74$   & 0.0091(02)   & 2000/09/01     & $375\pm 5$ & Asiago archive\\[0.3ex]
\hline 
\end{tabular}
\end{scriptsize}
\end{table*}

The full catalogue of spectra studied in this work is presented in 
Table~\ref{spectra}, complemented by additional information on the SN 
classifications and host-galaxy properties.

\subsection{Previously unpublished spectra}
\label{Previously unpublished spectra}

Our sample contains $26$ nebular spectra of $17$ SNe~Ib/c not previously 
published elsewhere. Another $4$ spectra of the Asiago archive were shown 
by \citet{Turatto03} and \citet{Valenti08} before. Most of these spectra 
were taken in the course of the ESO-Asiago SN monitoring programme in the 
1990s \citep{Turatto00a} using the ESO - La Silla 3.6m (equipped with 
EFOSC\,/\,EFOSC2), 2.2m (+ EFOSC2) and 1.5m (+ Boller \& Chivens spectrograph) 
Telescopes and the 1.54m Danish Telescope (equipped with DFOSC). From 2004 
onwards, several spectra were acquired through dedicated VLT programmes 
(VLT-U1 equipped with FORS2). The set is complemeted by single spectra taken 
with the Siding Spring 2.3m Telescope (+ double-beam spectrograph) and the 
Nordic Optical Telescope (+ ALFOSC). Details on the dates of the observations 
and the instrumental setup are summarised in Table~\ref{unpublished}. 

\addtocounter{table}{-1}
\begin{table*}
\caption[]{\textit{continued.} List of SNe~Ib/c included in the sample -- $2^{nd}$ part.} 
\center
\begin{scriptsize}
\begin{tabular}{llllccccl}
\hline \\[-1.7ex]
SN     & Type           & Host galaxy  & Morphology$^a$ & $v_\mathrm{rec}^b$ & Redshift & Date           & Epoch$^c$  & Reference \\[0.3ex]
\hline  \\[-1.7ex]
2000ew & Ic             & NGC 3810     & Sc             & $1049\pm114$   & 0.0035(04)   & 2001/03/17     & $112\pm12$ & Asiago archive\\
2002ap & BL-Ic          & NGC 628      & Sc             &  $657\pm 22^d$ & 0.0022(01)   & 2002/06/09     & $123\pm 1$ & \citealt{Foley03}\\
       &                &              &                &                &              & 2002/06/18     & $132\pm 1$ & \citealt{Foley03}\\
       &                &              &                &                &              & 2002/07/12     & $156\pm 1$ & \citealt{Foley03}\\
       &                &              &                &                &              & 2002/08/09     & $185\pm 1$ & \citealt{Foley03}\\
       &                &              &                &                &              & 2002/10/01     & $237\pm 1$ & \citealt{Foley03}\\
       &                &              &                &                &              & 2002/10/09     & $245\pm 1$ & \citealt{Foley03}\\
       &                &              &                &                &              & 2002/10/14     & $250\pm 1$ & SSO\,/\,Asiago archive\\
       &                &              &                &                &              & 2002/11/06     & $274\pm 1$ & \citealt{Foley03}\\
       &                &              &                &                &              & 2003/01/07     & $336\pm 1$ & \citealt{Foley03}\\
       &                &              &                &                &              & 2003/02/27     & $386\pm 1$ & \citealt{Foley03}\\
2003jd & BL-Ic          & MCG-01-59-021& SABm           & $5654\pm 78$   & 0.0188(03)   & 2004/09/11     & $317\pm 1$ & \citealt{Mazzali05}\\
       &                &              &                &                &              & 2004/10/18     & $354\pm 1$ & \citealt{Mazzali05}\\
2004aw & Ic             & NGC 3997     & SBb(M)         & $4900\pm118$   & 0.0163(04)   & 2004/11/14     & $236\pm 1$ & \citealt{Taubenberger06}\\
       &                &              &                &                &              & 2004/12/08     & $260\pm 1$ & \citealt{Taubenberger06}\\
       &                &              &                &                &              & 2005/05/11     & $413\pm 1$ & \citealt{Taubenberger06}\\
2004gt & Ic             & NGC 4038     & SBm(M)         & $1424\pm133$   & 0.0047(04)   & 2005/05/24     & $160\pm 5$ & Asiago archive\\
2005N  & Ib/c           & NGC 5420     & Sb             & $4885\pm198^d$ & 0.0163(07)   & 2005/01/21     &  $88\pm30$ & \citealt{Harutyunyan07}\\ 
2006F  & Ib             & NGC 935      & Sc(M)          & $4270\pm180$   & 0.0142(06)   & 2006/11/16     & $312\pm 7$ & \citealt{Maeda08}\\
2006T  & IIb            & NGC 3054     & SBb(R)         & $2560\pm182$   & 0.0085(06)   & 2007/02/18     & $371\pm 2$ & \citealt{Maeda08}\\
2006aj & BL-Ic          & anonymous    & late spiral    & $9845\pm250$   & 0.0328(08)   & 2006/09/19     & $204\pm 1$ & \citealt{Mazzali07a}\\
       &                &              &                &                &              & 2006/11/27     & $273\pm 1$ & MPA data base\\
       &                &              &                &                &              & 2006/12/19     & $295\pm 1$ & MPA data base\\
2006gi & Ic             & NGC 3147     & Sbc            & $2820\pm178^d$ & 0.0094(06)   & 2007/02/10     & $148\pm 5$ & Asiago archive\\
2006ld & Ib             & UGC 348      & SABd(R)        & $4168\pm 41$   & 0.0139(01)   & 2007/07/17     & $280\pm 4$ & MPA data base\\
       &                &              &                &                &              & 2007/08/06     & $300\pm 4$ & MPA data base\\
       &                &              &                &                &              & 2007/08/20     & $314\pm 4$ & MPA data base\\
2007C  & Ib             & NGC 4981     & SBbc(R)        & $1766\pm116$   & 0.0059(04)   & 2007/05/17     & $131\pm 4$ & MPA data base\\
       &                &              &                &                &              & 2007/06/20     & $165\pm 4$ & MPA data base\\
2007I  & BL-Ic          & anonymous    & late spiral    & $6445\pm250$   & 0.0215(08)   & 2007/06/18     & $165\pm 6$ & MPA data base\\
       &                &              &                &                &              & 2007/07/15     & $192\pm 6$ & MPA data base\\[0.3ex]
\hline 
\end{tabular}
\flushleft
$^a$ Classification according to LEDA (Lyon-Meudon Extragalactic Database, http:/$\!$/leda.univ-lyon1.fr/).\\
$^b$ Recession velocity in km\,s$^{-1}$, inferred from narrow H$\alpha$; error from `vmaxg' (LEDA).\\
$^c$ Epoch in days from $B$-band maximum light; photometry of SN~1985F from \citet{Tsvetkov86}.\\
$^d$ No H$\alpha$ visible; heliocentric host-galaxy recession velocity from NED (NASA/IPAC Extragalactic Database, http:/$\!$/nedwww.ipac.caltech.edu/) used.\\
\end{scriptsize}
\end{table*}

Applying the selection criteria mentioned above, our full sample consists of $39$ SNe 
with $98$ nebular spectra. It contains almost all suitable spectra up to the year $2004$ 
that could be retrieved from the literature, complemented by previously unpublished spectra 
from the Asiago Supernova Archive \citep[][http:/$\!$/web.oapd.inaf.it/supern/cat/]{Barbon93} 
and selected spectra obtained through dedicated programmes after $2004$. The observations 
thus span the entire era of SN CCD spectroscopy.

\begin{table}
\caption{Instrumental details of spectra from the Asiago archive and the MPA data base.} 
\label{unpublished}
\center
\begin{scriptsize}
\begin{tabular}{lll}
\hline \\[-1.7ex]
SN     & Date           & Instrumental setup                \\[0.3ex]
\hline \\[-1.7ex]
1990U  & 1990/11/23     & ESO\,3.6m + EFOSC + B300 + R300   \\
       & 1990/12/20     & ESO\,3.6m + EFOSC + B300          \\
1990W  & 1991/02/21     & ESO\,3.6m + EFOSC + B300 + R300   \\
       & 1991/04/21     & ESO\,3.6m + EFOSC + B300 + R300   \\
1990aj & 1991/01/29$^a$ & ESO\,2.2m + EFOSC2 + gr1          \\
       & 1991/02/22$^a$ & ESO\,3.6m + EFOSC + B300 + R300   \\
1996D  & 1996/09/10     & ESO\,1.5m + B\&C   + gt15         \\
1996aq & 1997/02/11     & ESO\,3.6m + EFOSC  + R300         \\
       & 1997/04/02$^b$ & ESO\,1.5m + B\&C   + gt15         \\
       & 1997/05/14     & ESO\,2.2m + EFOSC2 + gr1 + gr5    \\
1997B  & 1997/09/23     & ESO\,2.2m + EFOSC2 + gm5          \\
       & 1997/10/11     & Danish\,1.54m + DFOSC + gr5       \\
       & 1998/02/02$^a$ & ESO\,3.6m + EFOSC2 + gr6          \\
1997X  & 1997/05/13     & ESO\,2.2m + EFOSC2 + gr1 + gr5    \\
1997dq & 1998/05/30     & ESO\,3.6m + EFOSC2 + B300N + R300N\\
1999cn & 2000/04/08     & ESO\,3.6m + EFOSC2 + gr11         \\
1999dn & 2000/09/01     & ESO\,3.6m + EFOSC2 + gr12         \\
2000ew & 2001/03/17     & Danish\,1.54m + DFOSC + gm4       \\
2002ap & 2002/10/14     & SSO\,2.3m + DBS                   \\
2004gt & 2005/05/24     & VLT-U1 + FORS2 + 300V             \\
2006aj & 2006/11/27     & VLT-U1 + FORS2 + 300V + 300I      \\
       & 2006/12/19     & VLT-U1 + FORS2 + 300V             \\
2006gi & 2007/02/10     & NOT\,2.56m + ALFOSC + gm4         \\
2006ld & 2007/07/17     & VLT-U1 + FORS2 + 300V             \\
       & 2007/08/06     & VLT-U1 + FORS2 + 300V             \\
       & 2007/08/20     & VLT-U1 + FORS2 + 300V             \\
2007C  & 2007/05/17     & VLT-U1 + FORS2 + 300V             \\
       & 2007/06/20     & VLT-U1 + FORS2 + 300V + 300I      \\
2007I  & 2007/06/18     & VLT-U1 + FORS2 + 300V + 300I      \\
       & 2007/07/15     & VLT-U1 + FORS2 + 300V             \\[0.3ex]
\hline 
\end{tabular}
\flushleft
$^a$ Already shown by \citet{Turatto03}.\\
$^b$ Already shown by \citet{Valenti08}.\\
\end{scriptsize}
\end{table}

\begin{figure*}
   \centering
   \includegraphics[width=17.8cm]{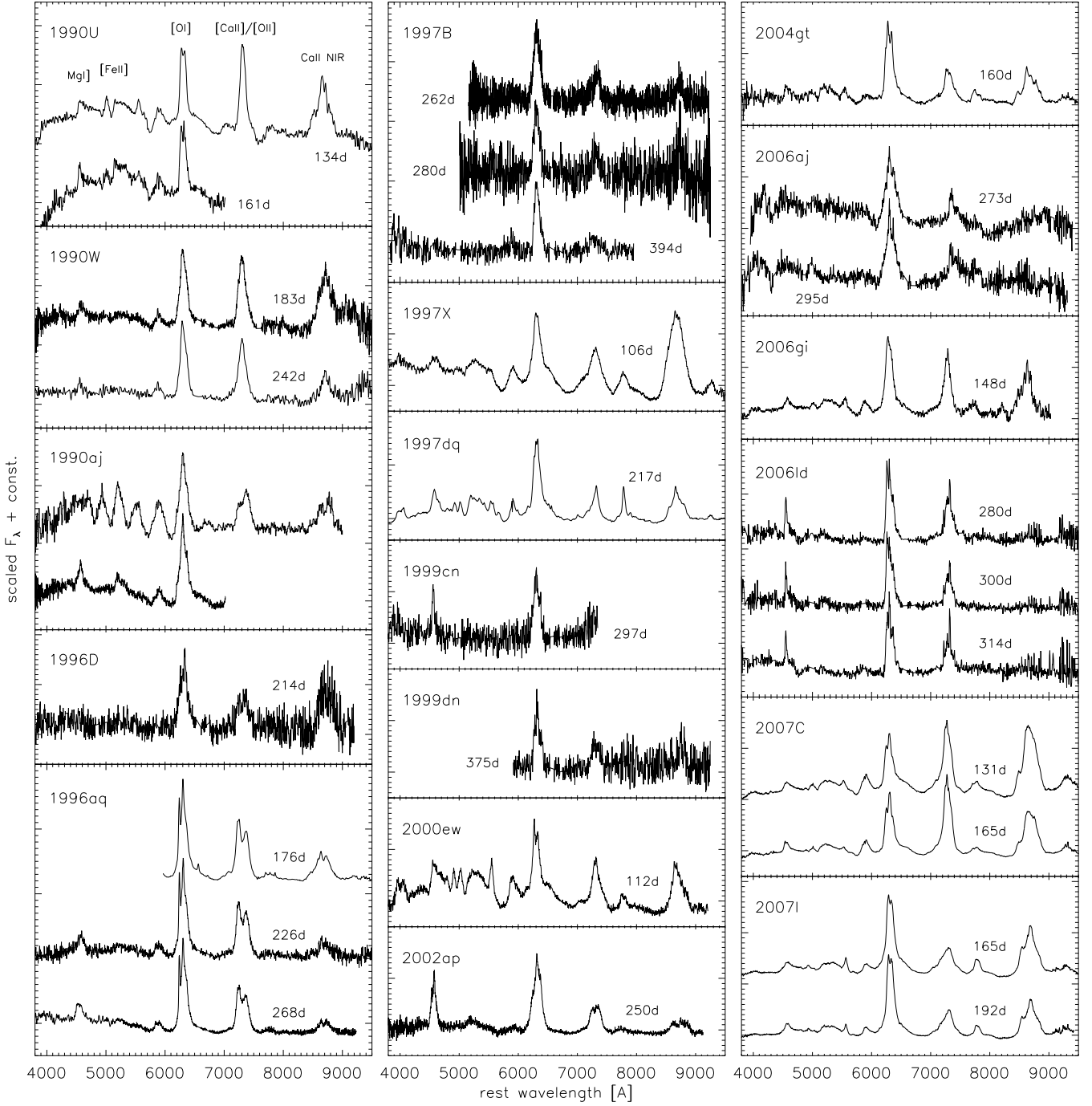}
   \caption[Spectra from the Asiago archive and the MPA data base, most of 
           them previously unpublished. All spectra are shown at their rest 
	   wavelength; overly strong continuum slopes and narrow host-galaxy 
	   emission features have been removed.]
	   {Spectra from the Asiago archive and the MPA data base, most of 
           them previously unpublished (cf. Table~\ref{unpublished}). All 
	   spectra are shown at their rest wavelength inferred (whenever 
	   possible) from narrow interstellar H$\alpha$ lines. Overly strong 
	   continuum slopes and narrow host-galaxy emission features have 
	   been removed for presentation purposes, and the spectra have been 
	   scaled and vertically displaced by arbitrary amounts. The major 
	   features are labelled in the first spectrum.}
   \label{fig:unpublished}
\end{figure*}

Spectra from the Asiago and MPA archives are presented in Fig.~\ref{fig:unpublished}.
They have been optimally extracted \citep{Horne86} using standard tasks in 
{\sc iraf}\footnote{{\sc iraf} is distributed by the National Optical Astronomy 
Observatories, which are operated by the Association of Universities for Research 
in Astronomy, Inc, under contract with the National Science Foundation.} or 
{\sc midas}, wavelength calibrated with respect to arc lamps, and flux calibrated 
using instrumental response curves obtained from spectrophotometric standard stars 
observed in the same nights. However, no attempt has been made to calibrate the 
fluxes to a proper absolute scale through a comparison to contemporaneous photometry. 

The spectra all show [O\,\textsc{i}] $\lambda\lambda6300,6364$ as one of 
their strongest features, complemented by other lines typical of SNe Ib/c at late 
phases, most notably [Ca\,\textsc{ii}] $\lambda\lambda7291,7323$\,/\,[O\,\textsc{ii}] 
$\lambda\lambda7320,7330$, Mg\,\textsc{i}] $\lambda4571$, a multitude of blended 
[Fe\,\textsc{ii}] lines around $5000$\,\AA, and the near-IR Ca\,\textsc{ii} triplet. 
In spectra taken less than $150$\,d after maximum a contribution from photospheric 
lines and a weak pseudo-continuum can be discerned. Some spectra also show evidence 
of an underlying stellar continuum caused by an imperfect host-galaxy subtraction.

\section{Fitting the oxygen line}
\label{Fitting the oxygen line}

[O\,\textsc{i}] $\lambda\lambda6300,6364$ is one of the dominant features of 
nebular spectra of stripped-envelope CC-SNe, and the most useful to study the 
ejecta geometry, in particular the degree of asphericity. Compared to the 
multitude of forbidden Fe lines found mostly at bluer wavelength and to the 
[Ca\,\textsc{ii}]\,/\,[O\,\textsc{ii}] feature around $7300$\,\AA, the 
[O\,\textsc{i}] $\lambda\lambda6300,6364$ doublet is largely isolated and 
unblended. In contrast to the weaker Mg\,\textsc{i}] $\lambda4571$ it lies 
in a region which is covered by almost all late-time spectra, and where the 
sensitivity of most spectrographs is at its maximum, allowing for relatively 
good S/N. Furthermore, oxygen is the most abundant element in the ejecta of 
stripped-envelope CC-SNe, thus tellingmore about the overall geometry than 
the distribution of a minor species like Ca.

In this section we describe our method to obtain constraints on the geometry 
of the ejected oxygen, which involves the characterisation of the expected 
line profile, the development of a suitable parametrisation, and the actual 
procedure applied to infer the best-fitting parameters for each spectrum.

\subsection{Line profiles}
\label{Expected line profiles}

In completely transparent SN ejecta the profile of a forbidden emission line traces 
the emissivity in this line, which in turn is determined by the spatial distributions 
and velocity fields of both the emitting species and $^{56}$Co, whose radioactive decay 
provides the energy to excite the line's upper level \citep{Fransson87,Fransson89}. 
In SNe~Ic the O- and Co-rich parts represent a significant fraction of the entire 
ejecta, and the \OIa\ feature traces a substantial amount of material. Thanks to 
homologous expansion ($r=vt$), the profile of the \OIa\ emission line is a 1D 
line-of-sight projection of the 3D oxygen emissivity distribution. 

In this work, we fit the oxygen feature with a Gaussian or -- for SNe with more complex 
line profiles -- with a superposition of multiple Gaussians. This provides at least 
qualitative information on the distribution (through the position and strength of the 
various components) and radial extent (through the components' FWHM) of excited oxygen 
in the SN ejecta. However, a full restoration of the 3D density distribution is not 
attempted, since the solution is highly degenerate.

\subsection{[O\,\textsc{I}] $\lambda6300$ and [O\,\textsc{I}] $\lambda6364$}
\label{O_i_6300 and O_i_6364}

As mentioned above, an advantage of the oxygen feature is its isolated position, 
unblended with lines from other elements. However, the feature itself is a 
doublet of [O\,\textsc{i}] $\lambda6300$ and [O\,\textsc{i}] $\lambda6364$, both 
forbidden M1 transitions which share the same upper level ($^3$P$_{1,2}$ -- 
$^1$D$_2$). The intensity ratio of these two lines depends on the ambient 
O\,\textsc{i} density, and can vary from 1\,:\,1 to 3\,:\,1 depending on the 
environmental conditions. The transition between the asymptotic values occurs at 
O\,\textsc{i} densities of $n(\textrm{O}\,\textsc{i}) \approx 10^{10}$ cm$^{-3}$. 
This fact has been theoretically derived by \citet{LiMcCray92} and \citet{Chugai92}, 
and observationally confirmed by \citet{PhillipsWilliams91} and 
\citet{SpyromilioPinto91} for SN~1987A, \citet{Leibundgut91} for SN~1986J, and 
\citet{Spyromilio91} for SN~1988A. 

Compared to the aforementioned SNe II, the stripped CC-SNe of our sample show 
larger ejecta velocities. While \citet{SpyromilioPinto91} measured a FWHM of 
$2800$ \kms for the \OIa\ line in SN~1987A, the SNe discussed here have FWHM 
of $\sim$\,$6000$ \kms$\!\!$. Assuming a Gaussian density profile with $6000$ 
\kms FWHM and $1 M_\odot$ of neutral oxygen homogeneously distributed within the 
ejecta, the central O\,\textsc{i} density would have dropped to $4$--$5 \times 
10^8$ cm$^{-3}$ by $100$\,d, an order of magnitude below the density where 
deviations from a 3\,:\,1 line ratio become apparent. Even if the oxygen was 
clumped on small scales within an overall Gaussian profile [as in SN~1987A, for 
which \citet{SpyromilioPinto91} suggested an oxygen filling factor of 
$\sim$\,$10\%$ based on the observed evolution of the \OIa\ ratio], ratios 
significantly different from 3\,:\,1 would only be encountered for very small 
filling factors ($< 10\%$). We therefore adopt the low-density limit (3\,:\,1) 
for all spectra, noting that possible small deviations at the earliest epochs 
do not severely affect any of our basic conclusions.

With this choice, it is sufficient to specify the amplitude, central wavelength 
and FWHM of the $\lambda6300$ line. The parameters for the $\lambda6364$ line are 
then forced. Hereafter we refer to such a set of three parameters as 
\textit{one component} of the fit to the [O\,\textsc{i}] profile; up to three such 
components were employed to obtain good fits.

\subsection{The fitting procedure}
\label{The fitting procedure}

The actual fitting was accomplished using the {\sc iraf} task {\sc nfit1D}, 
which is part of the {\sc stsdas} package. A user-defined fitting-function 
was introduced, which consisted of up to three components, as defined above.

The background level was determined by eye on both sides of the [O\,\textsc{i}] 
feature and interpolated linearly to account for a possible underlying continuum 
formed by residual photospheric lines or a contamination by star light. Since 
the background parameters are determined from a different wavelength region than 
the line parameters, the background fit is technically decoupled from the line 
fit, although of course the best-fit line parameters may be affected by the 
choice of the background. This can be problematic in relatively early spectra 
(up to $\sim$\,150\,d), where underlying photospheric lines are present, and the 
local continuum is more strongly inclined than at later times. However, tests 
with different choices of the continuum level have shown that the uncertainty 
introduced e.g. in the central wavelength of the line is $\la 5$\,\AA\ even in 
cases with very complex background. More typical uncertainties are $\la 2$\,\AA\ 
and thus smaller than the uncertainties in the wavelength calibration and 
redshift correction of most spectra.

For the up-to-nine parameters of the line fit (amplitude, position $\lambda$ 
and FWHM of up-to-three components) the main difficulty consisted in 
identifying the global minimum for a given number of components. Therefore, 
for fits with two or three components, a refined two-step fitting procedure 
was applied to avoid local minima. First, a set of synthetic line profiles 
were generated, changing all parameters by equidistant steps over a fairly 
wide range. The resulting profiles were compared to the observed ones, and 
the one with minimum RMS was identified. In a second step the values 
thus derived were used as initial guesses for {\sc nfit1D}, ensuring that 
the fit converged to the global minimum. We always started the fitting with 
one component. Other components were added only if the fit residuals strongly 
exceeded the noise level of the spectrum.

\begin{figure*}
   \centering
   \includegraphics[width=17.8cm]{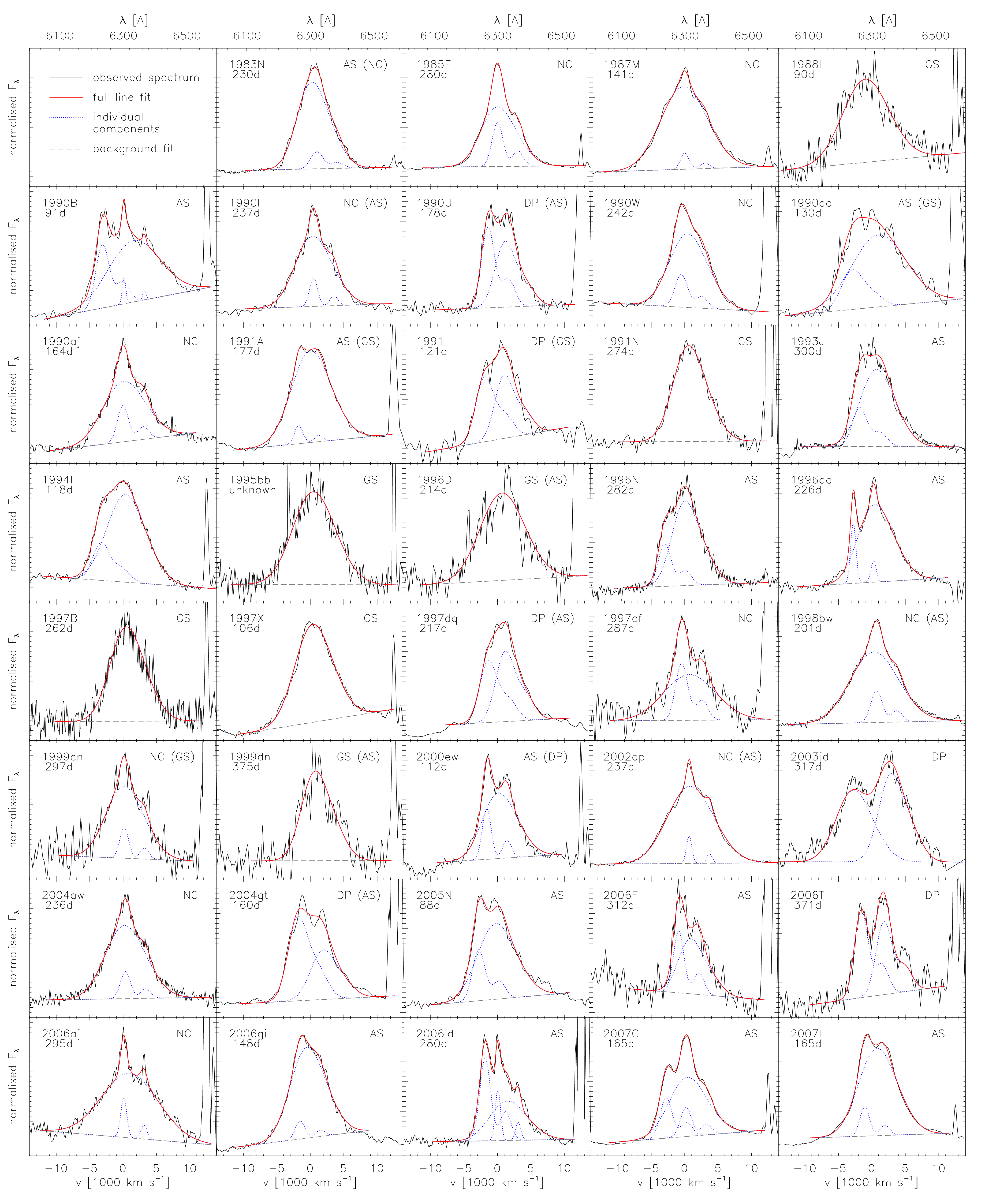}
   \caption[Observed \OIa\ $\lambda\lambda6300,6364$ features in the spectra of 
           our sample with multi-Gaussian fits overplotted. Individual components 
	   are shown, and the adopted linear background levels are indicated. The 
	   different categories as defined in Section~\ref{statistical analysis} 
	   are labelled as follows: GS for Gaussian, NC for narrow core, 
	   DP for double peak, AS for asymmetric\,/\,multi-peaked.]
	   {Observed \OIa\ $\lambda\lambda6300,6364$ features in the spectra of 
	   our sample (solid black lines) with multi-Gaussian fits 
	   (Tables~\ref{parameters12g} and \ref{parameters_3g}) overplotted 
	   (solid red lines). Individual components are overplotted as dotted blue 
	   lines, and the adopted linear background levels are indicated by dashed 
	   grey lines. H$\alpha$ was subtracted from the spectrum of SN~1993J (cf. 
	   Table~\ref{parameters12g}). The different line categories as defined in 
	   Section~\ref{classification scheme} are labelled as follows: GS for 
	   Gaussian, NC for narrow core, DP for double peak, AS for 
	   asymmetric\,/\,multi-peaked (alternative classifications are given in 
	   brackets). Only one spectrum of each object is included in the figure.}
   \label{fig:spectra_fitting}
\end{figure*}

The best-fitting values for single- and double-component fits to all spectra 
are reported in Appendix~\ref{Fit parameters} (Table~\ref{parameters12g}), 
while the parameters of three-component fits to individual spectra are given 
in Table~\ref{parameters_3g}. In these tables and during the further discussion, 
we define $\alpha_i$ as the integrated flux of the $i$-th component of the fit, 
normalised to the total integrated flux of the oxygen feature. The best fits 
are compared with the observed line profiles in Fig.~\ref{fig:spectra_fitting}.

\section{Blueshifted line centroids at early phases}
\label{blueshift}

\begin{figure}
   \centering
   \includegraphics[width=8.5cm]{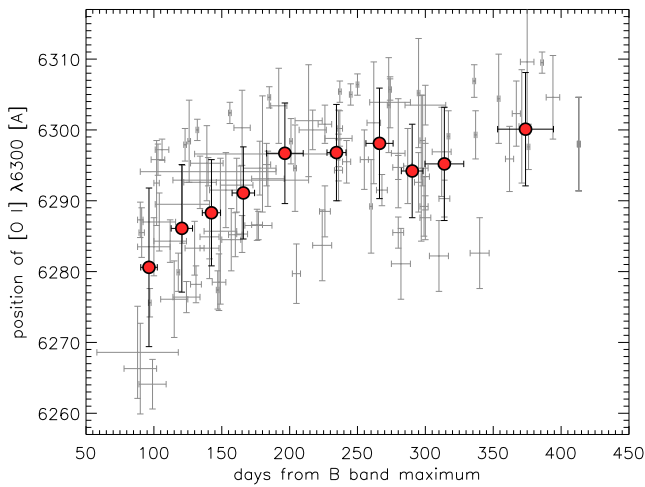}
   \caption[\OIa\ $\lambda6300$ line centroids, plotted vs. the epochs of the 
           spectra with respect to $B$-band maximum. A systematic blueshift 
	   can be discerned at epochs earlier than $200$\,d.]
	   {[O\,\textsc{i}] $\lambda6300$ line centroids (as inferred from the 
	   position of the $\lambda6300$ Gaussian in one-component fits), 
	   plotted vs. the epochs of the spectra with respect to $B$-band 
	   maximum. The filled circles represent bins of $10$ spectra. A 
	   systematic blueshift can be discerned at epochs earlier than 
	   $200$\,d.}
   \label{fig:blueshift}
\end{figure}

In Fig.~\ref{fig:blueshift} we plot the position of the $\lambda6300$-Gaussian 
in the one-component fits against the epoch of the spectra. The position should 
be a fair tracer of the actual line centroid. The scatter of the data points 
arises from the peculiarities of individual objects and uncertainties in their 
redshifts.

However, on top of this scatter Fig.~\ref{fig:blueshift} shows that there 
is a systematic trend of the [O\,\textsc{i}] feature being blueshifted in 
spectra taken earlier than $\sim$\,$200$\,d past maximum, and the effect is 
stronger the earlier the phase. In the spectra taken around $100$\,d, the 
average blueshift is $\sim$\,$20$\,\AA, corresponding to $\sim$\,$1000$ 
km\,s$^{-1}$. In the following, possible interpretations of the observed 
effect are discussed, and their suitability to explain the observations is 
considered.

(i) \textit{Ejecta geometry.} 
Line shifts such as those observed in [O\,\textsc{i}] could, in principle, 
arise from a one-sided ejecta geometry, caused for instance by low-mode 
convective instabilities \citep{Scheck04,Scheck06,Kifonidis06} or the 
Standing Accretion Shock Instability (SASI; \citealt{Blondin03}).
However, this can \textit{not} explain the decrease of the shifts with time. 
Moreover, there is no reason why particular ejecta geometries should result 
in a systematic \textit{blue}shift of the line centroid, as different 
spatial orientations should occur in the observed sample. Note that 
relativistic forward boosting is irrelevant at the observed ejecta velocities 
(no more than $\sim$\,$8000$ km\,s$^{-1}$, even in the extreme line wings).

(ii) \textit{Dust formation.}
As the SN ejecta expand and cool, the temperature eventually drops below the 
threshold where dust can form. Consequently, the light from the far side of 
the ejecta is partly absorbed, resulting in the suppression of the redshifted 
part of emission lines. This effect has been observed in some Type~II SNe, 
for instance SNe~1987A and 1999em, more than one year after explosion (cf. 
e.g. \citealt{Danziger89}, \citealt{Lucy89} and \citealt{Elmhamdi03}). In 
ordinary SNe~Ib/c dust formation has never been unambiguously detected at a 
few hundred days (e.g., \citealt*{Sollerman98}, \citealt{Matheson00b} and 
\citealt{Elmhamdi04}; but see also \citealt{Matthews02}). Moreover, dust 
formation (if present) should manifest itself in a line blueshift 
\textit{increasing} with time, the opposite of what we see in our sample, 
and hence cannot be a suitable explanation for our observations. 

\begin{figure}
   \centering
   \includegraphics[width=8.4cm]{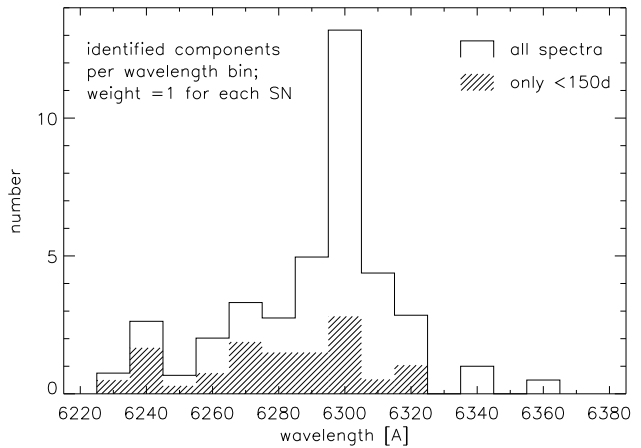}
   \caption[Histogram of components identified in the spectra with our Gaussian 
           fitting procedure. The empty histogram refers to the full sample, 
	   the shaded region to a subsample of spectra taken earlier than 
	   $150$\,d after maximum.]
	   {Histogram of components identified in the spectra with our Gaussian 
	   fitting procedure (wavelengths refer to [O\,\textsc{i}] $\lambda6300$; 
	   cf. Tables~\ref{parameters12g}, \ref{parameters_3g} and 
	   Fig.~\ref{fig:spectra_fitting}). To give equal weight to all SNe, the 
	   numbers have been rescaled such that every SN yields a contribution 
	   equivalent to one component (hence the fractional numbers). The empty 
	   histogram refers to the full sample, the shaded region to a subsample 
	   of spectra taken earlier than $150$\,d after maximum.}
   \label{fig:histogram}
\end{figure}

(iii) \textit{Contamination from other emission lines.}
In principle, other lines blended into the blue wing of [O\,\textsc{i}] 
$\lambda\lambda6300,6364$ could generate the observed blueshift of this feature. 
At early epochs, when a strong blueshift is observed, the contamination could 
e.g. arise from residual emission of permitted photospheric lines. The biggest 
problem with this idea is the apparent lack of suitable candidates. 
\citet{Elmhamdi04} suggested a contribution of Fe\,\textsc{ii} $\lambda6239$ 
in SN~1990I around $+90$\,d, but this is not expected to be a particularly 
strong line, and it is unclear why it should be so prominent while other, 
intrinsically stronger Fe lines are not. Moreover, a histogram of fit components 
(Fig.~\ref{fig:histogram}) does not show evidence of a distinct additional 
line at a wavelength shorter than $6300$\,\AA. Instead, in the subsample of 
spectra taken at $<$\,$150$\,d (shaded area in Fig.~\ref{fig:histogram}), 
the distribution of fit components smoothly smears out to shorter wavelengths. 
Finally, in Section~\ref{Mg profile} we will show that the profile of \MgIa\ 
$\lambda4571$ is similar to that of \OIa\ $\lambda6300$ in a majority of our 
spectra (also those with blueshifted lines), and an identical contamination 
in both lines is very unlikely.

\begin{figure}
   \centering
   \includegraphics[width=8.4cm,height=5.6cm]{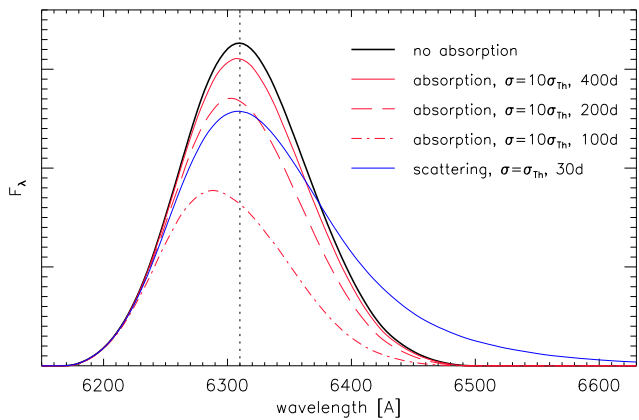}
   \caption[Synthetic profiles of \OIa\ $\lambda\lambda6300,6364$ calculated using 
           a Monte Carlo code.]
           {Synthetic profiles of \OIa\ $\lambda\lambda6300,6364$ calculated using a 
	   Monte Carlo code. A Gaussian density distribution, and both an emissivity 
           and opacity proportional to the density have been assumed. The pure-scattering 
           calculation is based on the Thomson cross-section $\sigma_\mathrm{Th}$ for 
           $e^-$-scattering, assuming on average singly ionised material. The cases of 
	   pure absorption were computed using a grey opacity with $\sigma = 10 
	   \sigma_\mathrm{Th}$. For comparison, the unabsorbed profile is also shown.}
   \label{fig:opaque_core}
\end{figure}

(iv) \textit{Opaque inner ejecta.}
The failure of other explanations and the characteristics of the fit-component 
histogram leave us with residual opacity in the core of the ejecta as a possible 
explanation for the observed blueshift \citep{Chugai92,WangHu94}. Optically thick 
inner ejecta could prevent light from the rear side of the SN from penetrating, 
creating a flux deficit in the redshifted part of emission lines. The opacity could 
be caused by e.g. densely packed weak Fe transitions (electron-scattering turns out 
to be at least an order of magnitude too weak). To see the effect on the line profile, 
we created a simple model using a Monte Carlo code (see Fig.~\ref{fig:opaque_core}), 
where photons are absorbed or scattered on their way to an observer with a 
probability proportional to the ambient matter density (grey opacity). 
A profile calculated for an unrealitically early epoch of $30$\,d assuming pure 
elastic electron-scattering shows a characteristic tail on the red side, but 
too little blueshift of the line core to be consistent with observations at $100$\,d 
(see Fig.~\ref{fig:opaque_core}). This demonstrates that Thomson scattering provides 
too little opacity, and, furthermore, would modify the line profile in an undesired 
way if it were strong enough. If, instead, the calculations are performed for grey 
absorption, good results can be obtained if the cross section is chosen appropriately. 
In particular, the observed time evolution of the line blueshift is reproduced 
qualitatively thanks to the $t^{-2}$ scaling of the column density.

\section{Statistical analysis, inferred ejecta geometries}
\label{statistical analysis}

The main intention of the multi-Gaussian line fitting is to derive information 
on the occurrence of different ejecta geometries in the sample. Of course, without 
additional assumptions it is not possible to restore the full 3D density 
distribution from its 1D projection given by the line profiles. While a forward 
calculation of emerging profiles for a given density distribution is staightforward, 
backward inference is highly degenerate.

A brief overview of some possible ejecta geometries and the corresponding observed 
line profiles is given in Table~\ref{correspondance}, illustrating that
in several cases the geometry cannot be determined with confidence. Yet, for most 
observed profiles we can at least exclude certain configurations. 

\begin{table*}
\caption{Selected oxygen geometries and corresponding line profiles.} 
\label{correspondance}
\center
\begin{footnotesize}
\begin{tabular}{lll}
\hline \\[-1.7ex]
Oxygen emissivity distribution       & Line profile                           & Global symmetry       \\[0.7ex]
\hline \\[-1.7ex]
radial Gaussian                      & Gaussian                               & spherically symmetric \\[0.4ex]
enhanced central density             & narrow core on top                     & spherically symmetric \\[0.4ex]
hard-edged homogeneous sphere        & parabolic                              & spherically symmetric \\[0.4ex]
spherical shell                      & flat-topped                            & spherically symmetric \\[1.2ex]
torus viewed from top                & narrow core                            & axisymmetric          \\[0.4ex]
torus viewed from the side           & double peak, symmetric to $\lambda_0$  & axisymmetric          \\[0.4ex]
torus viewed from intermediate angle & Gaussian-like                          & axisymmetric          \\[1.2ex]
small-scale clumpiness               & fine-structured peak                   & asymmetric            \\[0.4ex]
unipolar jet, one-sided blob         & extra-peaks\,/\,shoulders,             & asymmetric            \\
                                     & shifted with respect to $\lambda_0$    &                       \\[0.5ex]
\hline 
\end{tabular}
\end{footnotesize}
\end{table*}

\subsection{Taxonomy}
\label{classification scheme}

Guided by the results of the multi-Gaussian fitting, we introduce four principal 
classes of line profiles. Note that this classification scheme is a simplistic 
choice, based on experience acquired during the fitting. 

(i) \textit{Gaussian profiles (GS)}, well reproduced by single-component fits, 
with the residuals showing no evidence of a second component within the noise 
level. These profiles are expected from spherically symmetric ejecta with a 
nearly Gaussian emissivity distribution, but could alternatively be the outcome 
of e.g. axisymmetric explosions viewed from intermediate angles ($40$--$50$\degr, 
depending on the degree of asphericity; \citealt{Maeda08}).

(ii) \textit{Narrow-core lines (NC)}, characterised by an additional narrow 
component centred close to the rest wavelength \textit{and} atop the broad 
base of the line (offsets $\la 20$\,\AA). These are quite frequent and can be 
explained (a) by axisymmetric explosions with the emitting oxygen located in a 
torus or disk perpendicular to the line of sight (as inferred by \citealt{Mazzali01}, 
\citealt{Maeda02} and \citealt{Maeda06} for SN~1998bw), (b) by spherically symmetric 
ejecta with an enhanced core density, or (c) by a blob of oxygen moving nearly 
perpendicularly to the line of sight [cf. (iv)]. Note that the presence of dense 
cores has been suggested for several SNe to explain their line profiles and late-time 
light-curve slopes \citep{Iwamoto00,Mazzali00b,Maeda03,Mazzali07a,Mazzali07b}. 

(iii) \textit{Double-peaked profiles (DP)} with two comparably strong components, 
one blueshifted and the other redshifted by similar amounts. These are most 
readily explained by a torus-shaped oxygen distribution viewed nearly sideways 
(from angles of $\sim$\,$60$--$90$\degr\ to the symmetry axis; \citealt{Mazzali05} 
and \citealt{Maeda08}). No double peak can be realised in spherical symmetry. 
Hence, this class of line profile requires asphericity. The prototype of this 
class is SN~2003jd \citep{Mazzali05,Valenti08}.

(iv) \textit{Multi-peaked or asymmetric profiles (AS)}, produced by additional 
components of arbitrary width and shift with respect to the main component. 
These profiles are either indicative of ejecta with large-scale clumping, a 
single massive blob, or a unipolar jet. Like the double peaks, they cannot be 
reproduced within spherical symmetry.\\

Assigning our sample of SNe to these categories is sometimes ambiguous. For 
instance, an \OIa\ feature which consists of a main peak and a Doppler-shifted blob 
may appear double-horned, and can be confused with genuine double peaks formed by 
a toroidal oxygen distribution as defined in (iii). A further complication for the 
classification arises from possible bulk shifts of the [O\,\textsc{i}] feature with 
respect to its rest wavelength: as discussed in Section~\ref{blueshift} this does 
not necessarily have a geometric origin. 

\begin{figure*}
   \centering
   \includegraphics[width=12.0cm]{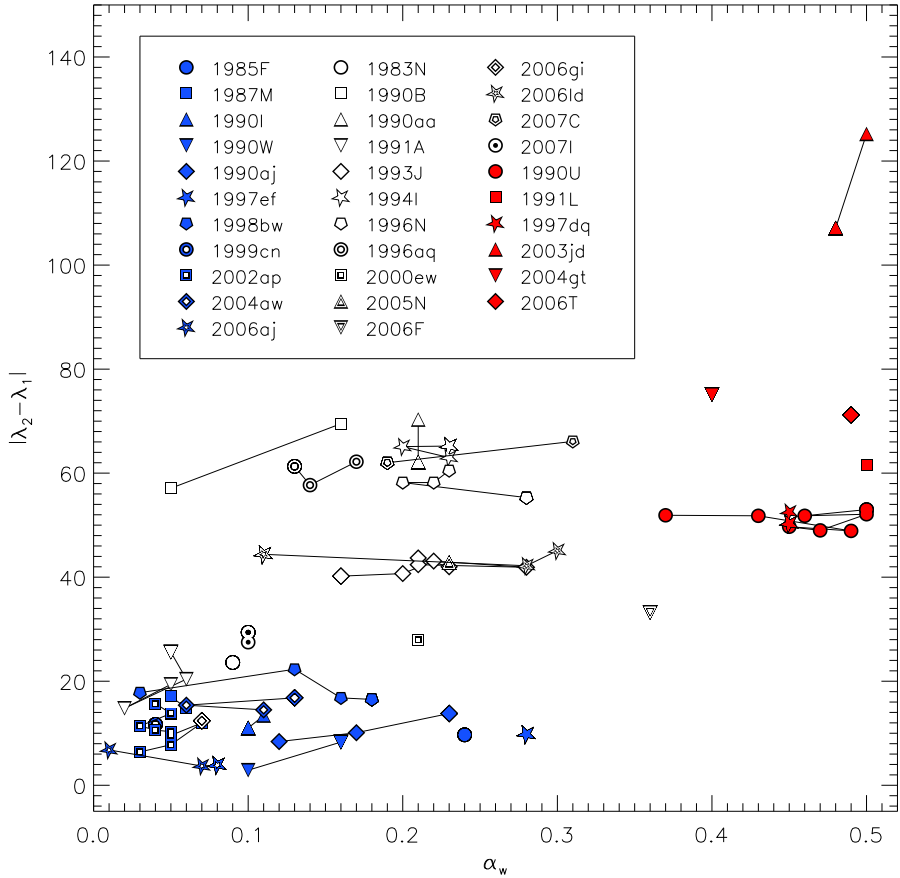}
   \caption[Absolute wavelength difference between two components found 
   by multi-Gaussian fitting of \OIa\ $\lambda6300$, as a function of 
   $\alpha_\mathrm{w}$, the relative flux of the weaker component. 
   Narrow-core SNe, double peaks and SNe with asymmetric or clumpy ejecta 
   appear fairly well separated in this plot.]
   {Absolute wavelength difference between two components found by 
   multi-Gaussian fitting of [O\,\textsc{i}] $\lambda6300$, as a function of 
   $\alpha_\mathrm{w}$, the relative flux of the weaker component. Objects 
   with single-Gaussian line profiles (cf. Fig.~\ref{fig:spectra_fitting}) 
   are not shown. Filled blue symbols stand for narrow-core SNe, filled red 
   symbols for double peaks, and open symbols for SNe with asymmetric or 
   clumpy ejecta. The different classes appear fairly well separated in 
   this plot.}
   \label{fig:classes}
\end{figure*}

Fig.~\ref{fig:classes} indicates the class membership of individual SNe. Only 
SNe fitted with two components [classes (ii) to (iv)] are included in this 
figure, which shows the absolute wavelength offset between the two fit components 
(Table~\ref{parameters12g}) as a function of $\alpha_\mathrm{w}$, the normalised 
flux of the weaker component. In this diagram, narrow-core SNe [class (ii)] 
populate a strip along the abscissa ($|\lambda_1 - \lambda_2| \la 20$\,\AA), 
double peaks [class (iii)] an area of larger wavelength offset and $0.4 \la 
\alpha_\mathrm{w} \leq 0.5$, while SNe with asymmetric or multi-peaked profiles 
[class (iv)] are mostly contained in a region characterised by $\alpha_\mathrm{w} 
\la 0.3$ and $|\lambda_1 - \lambda_2| > 20$\,\AA. Note, however, that some objects 
lying in the narrow-core strip actually belong to class (iv), since their weaker 
component has too large an offset ($> 20$\,\AA) from the rest wavelength to fulfil 
the criteria defined for class (ii). 

\subsection{Statistical evaluation}
\label{statistical evaluation}

In Table~\ref{statistics} the statistical summary of this analysis is 
presented. As we will see below, deviations from spherical symmetry affect 
all types of stripped-envelope CC-SNe, and are not reserved to particularly 
energetic or highly-stripped events. This is in agreement with the results 
of \citet{Modjaz08} and \citet{Maeda08}.

\begin{table}
\caption{SN taxonomy in terms of [O\,\textsc{i}] line profiles. The errors account for 
possible alternative classifications as indicated in Fig.~\ref{fig:spectra_fitting}.} 
\label{statistics}
\center
\begin{footnotesize}
\begin{tabular}{lll}
\hline \\[-1.7ex]
Category                   & Number                 & Percentage               \\[0.3ex]
\hline \\[-1.7ex]
(i) Gaussian               & \quad\ \,$7^{+4}_{-2}$ & \quad\ $18^{+10}_{- 5}$  \\
(ii) Narrow core           & \quad$11^{+1}_{-4}$    & \quad\ $28^{+ 3}_{-10}$  \\
(iii) Double peak          & \quad\ \,$6^{+1}_{-4}$ & \quad\ $15^{+ 3}_{-10}$  \\
(iv) Asymmetric\,/\,blobs  & \quad$15^{+8}_{-4}$    & \quad\ $39^{+21}_{-10}$  \\[0.3ex]
\hline 
\end{tabular}
\end{footnotesize}
\end{table}

\textit{Spherically symmetric objects.}
SNe whose [O\,\textsc{i}] profiles are well fit with single Gaussians make up 
little more than a quarter of all objects, even within the uncertainties. 
Considering all possibly spherical SNe [i.e., classes (i) and (ii)], and again 
including objects with ambiguous classification, we find their fraction to be 
just over $50\%$. Given that for some of these objects the S/N is too low to 
identify more than one component, that also jet-like explosions yield 
single-peaked symmetric profiles if viewed not too far from the jet axis, and 
that blobs moving roughly perpendicular to the line of sight can mimic narrow 
line cores, it is evident that this is really an upper limit for the number of 
spherical objects. This is exemplified by SN~1997X [class (i)], for which 
\citet{Wang01} measured exceptionally strong continuum polarisation, of the order 
of $4\%$, clearly indicating globally aspherical ejecta. However, at the same 
time the \OIa\ profile reveals no obvious deviation from spherical symmetry 
(Fig.~\ref{fig:spectra_fitting}). Hence, SN~1997X is probably intrinically 
aspherical, but viewed from a direction in which the line-of-sight projection 
of the oxygen emissivity mimics a spherical explosion. We thus speculate that 
probably more than half of all stripped-envelope CC-SNe are significantly aspherical. 

\textit{Symmetric double peaks.}
Between $5$ and $18\%$ of the SNe in our sample belong to class (iii), i.e. 
their line profiles are best reproduced by a symmetric double-peak configuration. 
This suggests a somewhat lower occurrance rate of double peaks than the samples 
of \citet{Maeda08} and \citet{Modjaz08}, where $28\%$ and $37\%$ of the SNe, 
respectively, were double-peaked. However, it should be noted that some of the 
double-peaked objects of \citet{Maeda08} and \citet{Modjaz08}, such as SNe~2004ao, 
2005aj and 2005bf, seem to lack symmetry about $\lambda_0$, and might have been 
placed in class (iv) in our scheme.

\textit{Jet-SNe.}
In the jet-models of \citet{Maeda06,Maeda08}, oxygen is distributed in a torus-like 
geometry perpendicular to the jet axis, and the \OIa\ profile is strongly 
viewing-angle dependent (see Table~\ref{correspondance}). These models  predict the 
ratio of narrow cores to double peaks to be $\sim$\,$1$\,:\,$5$, rather insensitive 
to the degree of asphericity. On the contrary, we find fewer double-peaked profiles 
than narrow cores, suggesting that the majority of these narrow cores do not 
originate from jets, but e.g. from an enhanced central density. This also means 
that only a rather small fraction of all SNe~Ib/c have a bipolar-jet geometry.
In fact, depending on whether moderately (BP2 of \citealt{Maeda06}) or strongly 
(BP8) aspherical models are used to evaluate the viewing-angle dependence of the 
profiles in detail, we find jet-SN fractions of $\sim$\,$50\%$ or $\sim$\,$25\%$ 
in our sample.

\begin{figure}
   \centering
   \includegraphics[width=8.0cm]{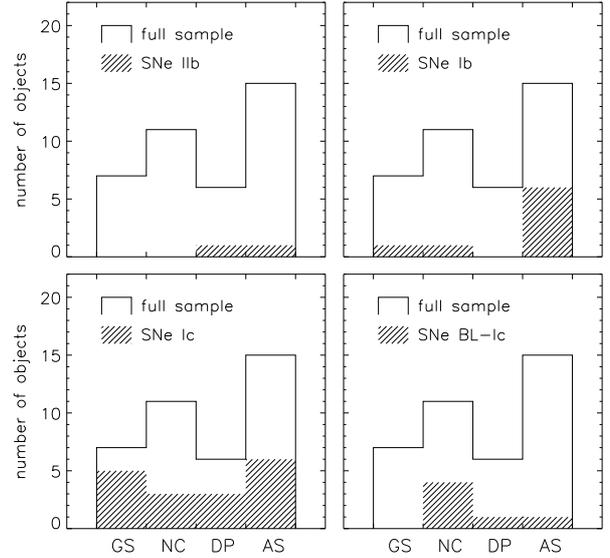}
   \caption[Allocation of the SNe in our sample to the line-profile classes 
           defined in Section~\ref{classification scheme}. SNe~IIb, SNe~Ib, SNe~Ic 
	   and BL-Ic SNe are separately shown in the different panels.]
	   {Allocation of the SNe in our sample to the line-profile classes 
	   defined in Section~\ref{classification scheme}: GS stands for 
	   `Gaussian' [class (i)], NC for `narrow core' [class (ii)], DP 
	   for `double-peaked' [class (iii)] and AS for `asymmetric\,/\,multi-peaked' 
	   [class (iv)] (cf. Table~\ref{statistics}). SNe~IIb, SNe~Ib, SNe~Ic and 
	   BL-Ic SNe are separately shown in the different panels.}
   \label{fig:types}
\end{figure}

\begin{figure}
   \centering
   \includegraphics[width=8.4cm]{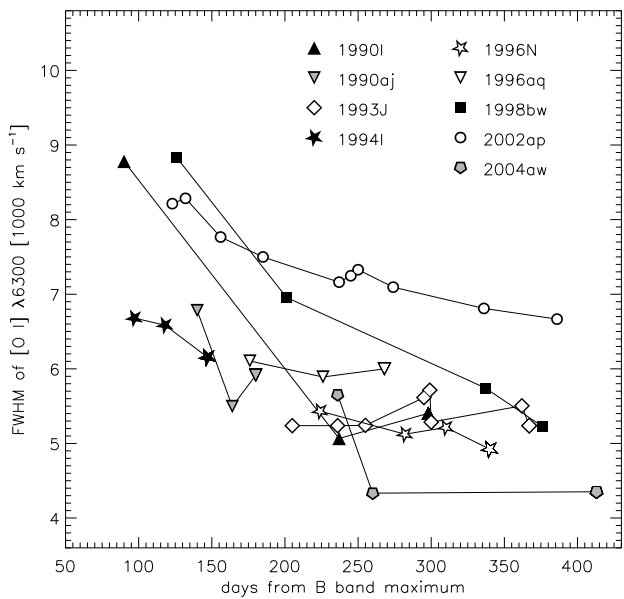}
   \caption[FWHM of the one-component Gaussian fit to the \OIa\ 
           $\lambda6300$ line as a function of time.]
	   {FWHM of the one-component Gaussian fit to the \OIa\ 
	   $\lambda6300$ line as a function of time. Only selected SNe 
	   with good temporal coverage are displayed.}
   \label{fig:fwhm}
\end{figure}

\textit{CC-SN subtypes.}
In Fig.~\ref{fig:types} we show how the traditional stripped-envelope CC-SN 
subtypes (i.e., BL-Ic, Ic, Ib and IIb) are distributed in terms of line-profile 
classes. Although the total number of objects is too small for robust statements, 
we are tempted to attribute some significance to the trends we can discern. 
While SNe~Ic, which form the majority of our sample ($59\%$), are relatively 
homogeneously distributed, SNe~Ib ($21\%$ of our sample) belong mainly to the 
multi-peaked\,/\,aspherical category. Finding objects with extended envelopes to 
show particularly strong aspericity appears counter-intuitive, and we have no 
convincing explanation for this behaviour. Similarly, broad-line SNe~Ic (BL-Ic; 
$\sim$\,$15\%$ of our sample) show a tendency towards nebular \OIa\ lines with 
narrow cores. At first, this seems to support the jet model. However, as discussed 
above, if all broad-line SNe~Ic with narrow core were interpreted as jet-events, 
a much larger number of double peaks would be expected.\footnote{Note that the 
BL-Ic SNe~1998bw and 2006aj, members of the narrow-core class, were discovered 
only after the detection of their associated GRBs. Given the particular geometry 
imposed by a GRB connection, a mild bias may be introduced into our statistics. 
Since we assert no claim to a strictly unbiased sample, we included these two 
objects in our analysis throughout this work.}
\newline
The FWHM of [O\,\textsc{i}] $\lambda6300$ (taken from the one-component fits), 
averaged over all spectra, increases from SNe~Ib/IIb ($5205 \pm 862\ \kms \!$) 
over normal-energetic SNe~Ic ($5942 \pm 1376\ \kms \!$) to broad-line SNe~Ic 
($7343 \pm 1724\ \kms \!$), reflecting the trend found in early-time spectra. 
A similar result has already been reported by \citet{Matheson01}, whose data set 
is included here. However, also the variation of the FWHM from object to object 
increases in this direction, indicating particularly strong diversity in the 
ejecta geometry of BL-SNe Ic. An observed trend towards smaller FWHM at later 
epochs (see Fig.~\ref{fig:fwhm}) could be explained by changes in the excitation 
conditions as a consequence of decreasing densities, such as a transition to more 
local positron deposition as the dominant excitation mechanism.

\section{Discussion of individual objects}
\label{Discussion of individual objects}

In the previous sections we have presented simple Gaussian fits to the 
[O\,\textsc{i}] $\lambda\lambda6300,6364$ features in nebular spectra of $39$ 
stripped-envelope CC-SNe, and found substantial diversity in the line profiles 
with deviations from spherical symmetry in a majority of the objects. However, 
remarkable patterns can also be discerned within the zoo of line profiles. In 
the following the properties of some interesting individual SNe or groups of 
objects are discussed in more detail.

\subsection{SNe~1994I, 1996N and 1996aq: Doppler-shifted blobs and neutron-star kicks?}
\label{1994I}

\citet{Sollerman98} noticed a blueshift of [O\,\textsc{i}] 
$\lambda\lambda6300,6364$ in late-time spectra of SN~1996N. We confirm this 
result, obtaining an overall blueshift of $850$ \kms\ in one-component fits. 
The epochs of the spectra are too late ($\geq$\,$180$\,d) to explain this with 
optical-depth effects. In a two-component setup, which yields a much better fit 
to the asymmetric line profile, the main component is nearly at rest, but the 
second one is blueshifted by $\sim$\,$3000$ \kms (Fig.~\ref{fig:spectra_fitting} 
and Table~\ref{parameters12g}), prominent and broad (FWHM $=2300$ km\,s$^{-1}$, 
$\alpha$\,$\sim$\,$0.2$ to $0.3$).

Here we stress the similarity of the [O\,\textsc{i}] line profiles of SNe~1996N 
and 1994I (although the latter was only observed at earlier phases, not later 
than $150$\,d past maximum). Finding such an unusual [O\,\textsc{i}] profile in 
SN~1994I was unexpected, given that this is one of the most soundly studied 
SNe~Ic to date, and, to our knowledge, this fact has never been commented on in 
the literature. SN~1996aq features an apparently different, double-peaked 
[O\,\textsc{i}] line. However, the fitting suggests that the main difference 
is the width of the blueshifted component, which is significantly narrower in 
SN~1996aq (FWHM $=1000$ km\,s$^{-1}$). This results in a separation of the 
$\lambda6300$ and $\lambda6364$ lines and thus a two-horned appearance.

The most convincing explanation for the observed line profiles in these three 
SNe is provided by blobs moving towards the observer at high velocity. If such 
a blob has an average composition and no enhanced excitation (as would be the 
result of an enhanced Co abundance), it has to carry substantial mass to 
account for the observed emission. In fact, in this simple scenario the mass 
fraction would be given by the fractional flux of the clump, $\alpha_\mathrm{w}$. 
In SN~1996aq, $\alpha_\mathrm{w}$ is about $0.15$, in SNe~1994I and 1996N 
$0.2$--$0.3$. Moving at a velocity $\ga 3000$ \kms$\!\!$, the blob carries 
enormous momentum, which has to be counter-balanced for the sake of momentum 
conservation. Since the remainder of the oxygen emission is centred at rest, 
the compensation has to be provided by Fe- or Si-rich material, or the compact 
remnant of the core collapse. Strong neutron star kicks (up to several hundred 
\kms) have indeed been observed \citep*[e.g.][]{Cordes93} and reproduced in 
simulations of anisotropic explosions with dominant dipole ($l=1$) mode in the 
ejecta \citep{Burrows96,Scheck04,Scheck06,Burrows07}. 
To estimate the kick velocities consistent with our measurements, we make a 
simple calculation for SN~1994I. With a total ejecta mass of $1.2\,M_\odot$ 
\citep{Sauer06} and an $\alpha_\mathrm{w}$ of $0.22$, the blueshifted blob contains 
$0.26\,M_\odot$ if a homogeneous composition is assumed throughout the ejecta. To 
compensate the momentum, a typical neutron star of $1.3\,M_\odot$ needs a kick 
velocity of $\sim$\,$600$ \kms$\!\!$ if moving along the line of sight, which is 
not an unreasonable number.

\subsection{SNe~1998bw and 2002ap: narrow cores?}
\label{1998bw}

SNe~1998bw and 2002ap share a similar [O\,\textsc{i}] $\lambda\lambda6300,6364$ 
line profile with a narrow component on top of a much broader base. In SN~1998bw, 
this was attributed to emission from a disk- or torus-shaped oxygen distribution 
viewed nearly from the top \citep{Mazzali01,Maeda02,Maeda06}. Together with Fe 
lines being broader than [O\,\textsc{i}] $\lambda\lambda6300,6364$, this gave rise 
to the idea of a strongly aspherical, jet-like explosion. For the nebular spectra 
of SN~2002ap, \citet{Foley03} remarked on a similarity of the line profiles with 
those of SN~1998bw, and \citet{Mazzali07a} suggested asphericity here as well, 
although the narrow peak might have been caused by a dense core in the ejecta.

\begin{figure}
   \centering
   \includegraphics[width=8.4cm]{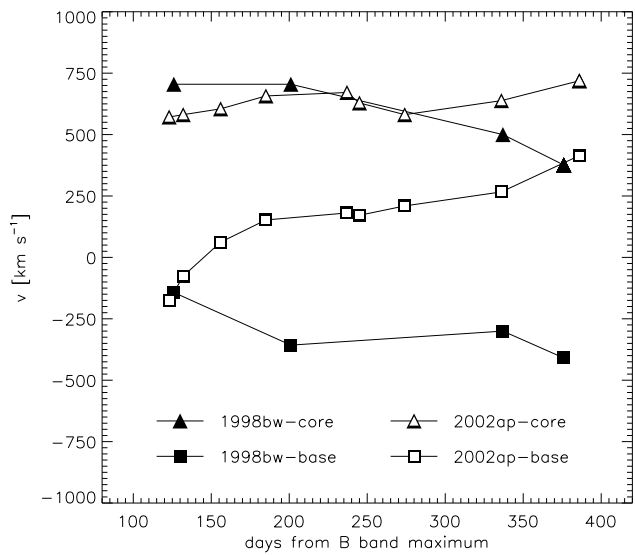}
   \caption[Shifts of the broad line bases and the narrow cores of \OIa\ 
           $\lambda6300$ in SNe~1998bw and 2002ap as a function of time.]
	   {Shifts of the broad line bases and the narrow cores of \OIa\ 
	   $\lambda6300$ in SNe~1998bw and 2002ap as a function of time 
	   (cf. Table~\ref{parameters12g}).}
   \label{fig:NC}
\end{figure}

However, in both SNe the narrow components are redshifted with respect to both the 
broad base and the rest wavelength $\lambda_0$ \citep{Patat01,Leonard02,Foley03}. 
From the best-fit parameters reported in Table~\ref{parameters12g}, mean redshifts 
of $586 \pm 162$ km\,s$^{-1}$ and $657 \pm 62$ km\,s$^{-1}$ with respect to $\lambda_0$ 
are inferred for SNe~1998bw and 2002ap, respectively (cf. Fig~\ref{fig:NC}). Relative 
to the broad bases, mean offsets of $876 \pm 129$ km\,s$^{-1}$ and $500 \pm 138$ 
km\,s$^{-1}$ are observed. 
Such large offsets (comparable to the narrow component's FWHM) are not expected 
for models with enhanced central density or the bipolar-jet scenario favoured for 
SN~1998bw. Instead, the nebular \OIa\ lines should be symmetric and centred at their 
rest wavelength unless the putative jet was rather one-sided. In that case, the lack 
of symmetry between the two hemispheres would explain the observed \OIa\ line shifts. 
In this work, we conservatively classify SNe~1998bw and 2002ap as \textit{possible} 
narrow cores (Fig.~\ref{fig:spectra_fitting}), considering the presence of a blob 
another possibility. Other SNe of the narrow-core group exhibit similar line offsets 
(cf. Table~\ref{parameters12g}), but mostly less pronounced than in SNe~1998bw and 
2002ap.

Inspired by the modelling of \citet{Tomita06} and \citet{Mazzali07a}, we examined 
an alternative, possibly more physical configuration for SN~2002ap. This consisted 
of a jet (or, more generically, a bipolar explosion) viewed strongly off-axis, but 
with an additional density enhancement at low velocity. To test the consistency with 
the observed \OIa\ line profile, we employed a three-component fit consisting of a 
strictly symmetric double peak plus a central narrow component ($7$ effective 
parameters; see Table~\ref{parameters_3g}), and compared with the 'broad peak + 
narrow core' setup adopted for SN~2002ap throughout the rest of this paper. 

It turns out that the two configurations perform similarly well, the only advantage 
of the two-component fit being that one fewer parameter is involved. It is therefore 
difficult to decide on the basis of these fits which ejecta geometry (on-axis jet, 
spherical ejecta\,+\,dense core, spherical ejecta\,+\,blob or off-axis jet\,+\,dense 
core) is most likely. In the 'DP + NC' configuration, about $15\,\%$ of the emitting 
mass would be contained in the dense core. This is in good agreement with the size 
of the core that was added to reproduce the late-time light curve of SN~2002ap in 
the models ($0.5$ out of a total of $3.0\,M_\odot$; \citealt{Tomita06}). The good 
quality of the fit and the consistency with sophisticated modelling make the 'DP + 
NC' scenario an attractive possibility for SN~2002ap.\footnote{Also other SNe of the 
narrow-core class can be adequately fit with three components in a configuration 
similar as in SN~2002ap. This could help to solve the 'problem of missing double 
peaks', see Section~\ref{statistical evaluation}). However, for the GRB-related 
SNe~1998bw and 2006aj a 'DP + NC' configuration -- though providing a good fit (cf. 
Table~\ref{parameters_3g}) -- is not expected to be correct, since these SNe are 
supposed to be viewed along the jet axis. Lacking modelling predictions for most 
other SNe, we have not explored the 'DP + NC' option any further.}

\subsection{SNe~2003jd, 2000ew, 2004gt and 2006T: genuine double peaks and impostors}
\label{2003jd}

Besides SNe~2003jd and 2006T, the prototypes of the double-peaked class, up to 
$5$ other SNe have \OIa\ lines that agree with a DP configuration. The problem 
with most of these objects is that the separation of the two fit components is 
smaller than in SNe~2003jd and 2006T, and often similar to the separation of the 
two \OIa\ lines in the doublet. This leaves room for alternative interpretations, 
in particular the possibility that the two horns observed e.g. in SNe~2000ew and 
2004gt may actually originate from the deblended $\lambda6300$ and $\lambda6364$ 
lines of a single narrow, blueshifted component. In fact, in SN~2000ew the line 
profile is better reproduced by the latter configuration. This is the reason for 
its primary association with the AS class (see Fig.~\ref{fig:spectra_fitting}).

In SNe~2000ew and 2004gt also the intensity ratio of the two peaks appears inverted 
with respect to those of SNe~2003jd and 2006T, the blue peak being stronger than 
the red one. In a classical DP configuration (cf. Section~\ref{classification scheme}), 
however, the red peak will always be stronger since, for the given separation 
of the peaks, the $\lambda6364$-line of the blue component blends with the 
$\lambda6300$-line of the red one. To circumvent this problem, a more complex 
ejecta structure with additional blueshifted emission on top of an otherwise 
symmetric profile may be assumed. This yields a good fit for SN~2004gt, cf. 
Table~\ref{parameters_3g}. Alternatively, the original toroidal oxygen distribution 
may be unchanged, but the redshifted emission component is damped owing to 
optically thick inner ejecta (cf. Section~\ref{blueshift}). Since the spectra 
of SNe~2000ew and 2004gt are both relatively young ($112$ and $160$\,d, 
respectively) compared to those of SNe~2003jd and 2006T (cf. Table~\ref{spectra}), 
this may indeed be a possibility. 

Due to its ejecta velocities and its double-peaked \OIa\ profile, SN~2003jd 
has been proposed to be associated with a GRB viewed strongly off-axis 
\citep{Mazzali05}. In that case, the $\gamma$- and X-ray emission of the 
GRB would not be seen because of the strong collimation of the jet. However, 
depending on the jet propagation model a radio afterglow would possibly be 
observable at late phases. SN~2003jd was not detected at radio wavelengths 
\citep{Soderberg06}, so that its association with a GRB remains uncertain. 
The second object with a similar \OIa\ line profile, SN~2006T, was classified 
as SN IIb \citep{Blondin06}. This makes it \textit{per se} a poor candidate 
for a GRB-SN, since the relativistic jet of a GRB would have to penetrate the 
He and H shells and probably die before reaching the surface. This supports 
the view of \citet{Modjaz08} and \citet{Maeda08} that strong asphericity is 
ubiquitous in core-collapse SNe, and not necessarily a signature of an 
association with a GRB (see also Section~\ref{statistical evaluation}).

\section{The profile of M\lowercase{g}\,{\sc i}] $\lambda4571$}
\label{Mg profile}

Hydrodynamic explosion models \citep{Maeda06} suggest that Mg and O should 
have similar spatial distributions within the SN ejecta, which may deviate 
significantly from those of heavier elements such as Fe or Ca (see also 
\citealt{Mazzali05}). This should result in the profiles of isolated Mg and 
O emission lines being similar, which so far has been shown to hold for 
individual SNe \citep{Spyromilio94,Foley03}. Here we test this for our 
entire sample, examining the profile of the semi-forbidden Mg\,\textsc{i}] 
$\lambda4571$ line (3$s^2$~$^1$S$_0$ -- 3$s$3$p$~$^3$P\degr$\!_1$) in all 
spectra with sufficient S/N in the wavelength range of interest.

\begin{figure}
   \centering
   \includegraphics[width=8.4cm]{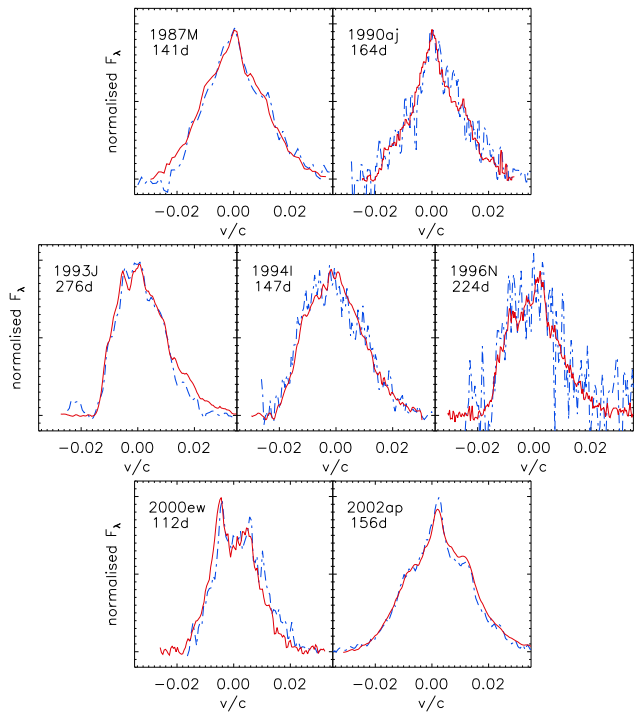}
   \caption[Comparison of \MgIa\ $\lambda4571$ and \OIa\ $\lambda\lambda6300,6364$ 
           line profiles (with a second component added to \MgIa\ artificially 
	   to account for the doublet nature of \OIa). A subsample of objects 
	   with a good overlap is shown.]
	   {Comparison of Mg\,\textsc{i}] $\lambda4571$ and [O\,\textsc{i}] 
	   $\lambda\lambda6300,6364$ line profiles (with a second component 
	   added to Mg\,\textsc{i}] artificially to account for the doublet 
	   nature of [O\,\textsc{i}], see discussion). A subsample of objects 
	   with a good overlap is shown. The dot-dashed blue line is the 
	   modified \MgIa, the solid red line \OIa.}
   \label{fig:Mg_vs_O_1}
\end{figure}

\begin{figure}
   \centering
   \includegraphics[width=8.4cm]{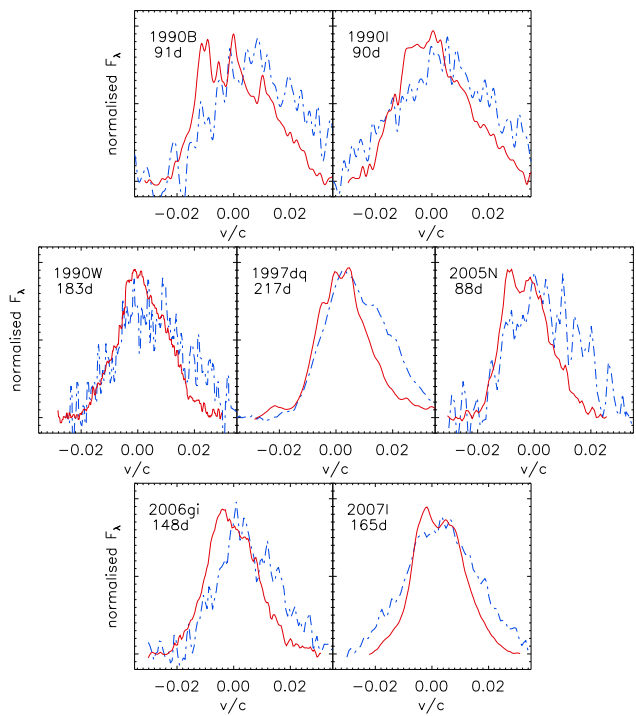}
   \caption[The same as Fig.~\ref{fig:Mg_vs_O_1}, but showing spectra with 
           evident differences in the \OIa\ $\lambda\lambda6300,6364$ and the 
	   modified \MgIa\ $\lambda4571$ profiles.]
	   {The same as Fig.~\ref{fig:Mg_vs_O_1}, but showing spectra with 
	   evident differences in the [O\,\textsc{i}] $\lambda\lambda6300,6364$ 
	   and the modified Mg\,\textsc{i}] $\lambda4571$ profiles.}
   \label{fig:Mg_vs_O_2}
\end{figure}

A direct comparison of the Mg\,\textsc{i}] and [O\,\textsc{i}] lines is 
hindered by the fact that, unlike Mg\,\textsc{i}], the [O\,\textsc{i}] feature 
is a doublet. However, as described in Section~\ref{O_i_6300 and O_i_6364}, we 
assumed that in our nebular SN~Ib/c spectra the $\lambda6300$ and $\lambda6364$ 
lines have a ratio of $3:1$. Therefore, to compensate we first isolated the 
Mg\,\textsc{i}] $\lambda4571$ feature, subtracting a linearly fit background. 
Then we rescaled the \MgIa\ line to 1/3 of its initial intensity, shifted it 
by $46$\,\AA\ (equivalent to the $64$\,\AA\ offset of the two \OIa\ lines) 
and added it to the original profile. This modified \MgIa\ profile can then 
be compared with the observed \OIa\ feature. 

\begin{figure}
   \centering
   \includegraphics[width=8.4cm]{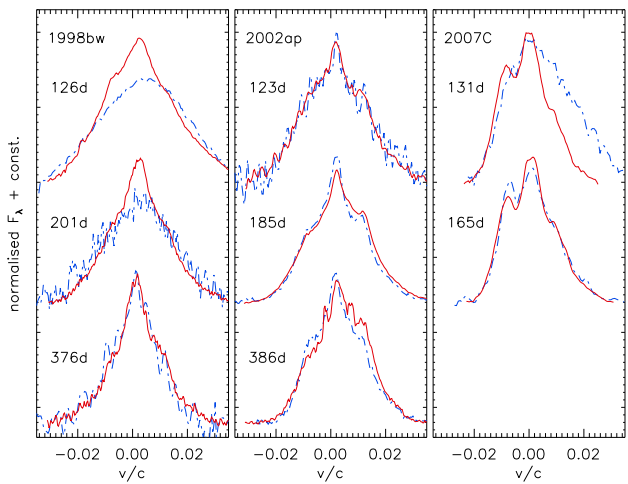}
   \caption[Evolution of the \OIa\ $\lambda\lambda6300,6364$ and the modified 
           \MgIa\ $\lambda4571$ line profiles of SNe~1998bw, 2002ap and 2007C 
	   with phase. While in SN~2002ap the profiles are always similar, 
	   significant differences are apparent in the earlier spectra of 
	   SNe~1998bw and 2007C.]
	   {Evolution of the [O\,\textsc{i}] $\lambda\lambda6300,6364$ (solid 
	   red) and the modified Mg\,\textsc{i}] $\lambda4571$ (dot-dashed blue) 
	   line profiles of SNe~1998bw, 2002ap and 2007C with phase. While in 
	   SN~2002ap the profiles are always similar, significant differences 
	   are apparent in the earlier spectra of SNe~1998bw and 2007C.}
   \label{fig:Mg_evolution}
\end{figure}

For most objects of our sample ($\sim$\,$65\%$), even those with rather 
complex ejecta geometry, we find an impressive similarity of the \MgIa\ and 
\OIa\ line profiles within the noise level and the uncertainty in subtracting 
the background (see Fig.~\ref{fig:Mg_vs_O_1}). This indicates that the spatial 
distribution of Mg and O in the ejecta is generally similar. However, 
there are some noticeable exceptions with a poor match of the [O\,\textsc{i}] 
and modified Mg\,\textsc{i}] features. Examples are shown in 
Fig.~\ref{fig:Mg_vs_O_2}. Most of these spectra are not very late, typically 
$< 200$\,d, which leaves room for the following explanations: 

(i) In the affected objects, the O- and Mg-rich parts of the ejecta may indeed 
have different geometry owing to the chemical stratification of the progenitor 
star and the hydrodynamics of the explosion. However, within this scenario 
differences in the line profiles should persist during the entire nebular phase. 
In SN~1998bw, for which a late nebular spectrum ($376$\,d) is available, the 
differences visible at earlier epochs ($\la 200$\,d) are observed to 
vanish with time (Fig.~\ref{fig:Mg_evolution}). SN~2007C undergoes a similar 
evolution even more rapidly.

(ii) Alternatively, the \MgIa\ or the \OIa\ features may be contaminated by 
nebular emission lines of other elements. For the \OIa\ feature the possibility 
of a contamination on the blue side has been discussed in Section~\ref{blueshift}, 
and found to be unlikely. Also, the fact that in most of the affected spectra the 
Mg\,\textsc{i}] line is broader than the [O\,\textsc{i}] line and changes more 
strongly with time (cf. Figs.~\ref{fig:Mg_vs_O_2} and \ref{fig:Mg_evolution}) 
suggests that the \MgIa\ rather than the \OIa\ line may be blended with other 
lines at early epochs.

(iii) The Mg\,\textsc{i}] $\lambda4571$ line is located in a region shaped by 
many strong Fe\,\textsc{ii} features during the photospheric phase. Hence, it 
is possible that the emission peak tentatively identified as \MgIa\ is mostly 
produced by underlying photospheric Fe lines in some earlier spectra. This 
would not only explain the observed evolution in SN~1998bw (where Fe features 
are particularly strong and persistent, see e.g. \citealt{Mazzali05}), but also 
the better agreement of \OIa\ and \MgIa\ in SN~2002ap early on: SN~2002ap 
features the strongest \MgIa\ $\lambda4571$ line ever observed \citep{Foley03} 
but only weak Fe, and it is hence not unexpected that \MgIa\ quickly dominates 
over photospheric residuals (Fig.~\ref{fig:Mg_evolution}).

\section{Conclusions}
\label{Conclusions}

We have studied the profiles of nebular emission lines in stripped-envelope 
CC-SNe, with the aim of constraining the ejecta morphology, and in particular 
the degree of asphericity of the explosions. The study was based on $98$ nebular 
spectra of $39$ different SNe, some of which were not published before. The size 
of the sample gives it statistical significance.
We have concentrated on the profile of [O\,\textsc{i}] $\lambda\lambda6300,6364$, 
since this is usually the strongest feature in nebular SN~Ib/c spectra, and not 
severely contaminated by other lines. We performed a multi-parameter Gauss-fitting 
of this feature in all our spectra, with the position, FWHM and intensity of the 
$\lambda6300$ Gaussian being free parameters (and the $\lambda6364$ line duly 
added with fixed offset and intensity ratio of 1/3). 
Using this approach, a variable number of emission components, their widths and 
Doppler shifts could be identified. Compared to spectral modelling this method 
has the advantage of being fast, capable of dealing with complex profiles, and 
independent of an accurate flux calibration.

Despite the large variety in line profiles encountered, we can divide the SNe 
into four morphologically different groups on the basis of the best-fit parameters: 
SNe with simple Gaussian line profiles ($\sim$\,$18\%$ of all objects), SNe 
with narrow line cores atop broad bases ($\sim$\,$28\%$), objects with symmetric 
double peaks ($\sim$\,$15\%$), and objects with evidence for blobs or overall 
asymmetric line profiles ($\sim$\,$39\%$). Since this classification refers to 
the structure inferred from the fitting procedure, we believe it to be more 
relevant for the true ejecta geometry than a pure visual inspection of the 
line. The results of our analysis suggest that probably at least half of all 
SNe Ib/c are aspherical. The fraction of double peaks is too small for a jet to 
be a ubiquitous feature in a majority of the objects. Even among narrow-core SNe, 
only a small fraction of the objects may have a jet-like ejecta morphology. 
Instead, a central density enhancement appears to be a likely solution for many 
members of this class. A majority of broad-line SNe~Ic have narrow line cores, 
whereas most SNe~Ib exhibit asymmetric or multi-peaked line profiles. 

Bulk shifts of the \OIa\ feature and strongly Doppler-shifted, massive blobs are 
observed in some objects, and expected to be the signature of very one-sided 
explosions. If momentum conservation is provided by neutron-star kicks, the 
inferred kick velocities are compatible with those of the fastest-moving 
neutron-stars in the Galaxy.

In spectra taken earlier than $\sim$\,$200$\,d after maximum light, a systematic 
blueshift of the [O\,\textsc{i}] $\lambda\lambda6300,6364$ line centroids can be 
discerned, becoming more pronounced with decreasing phase. Geometrical effects and 
dust formation within the ejecta seem to be excluded as origins of the blueshift. 
Contamination from other elements may play a role in some, but not all SNe. Hence, 
residual opacity in the inner ejecta remains the most likely explanation for the 
observed shift, as photons emitted on the rear side of the SN are scattered or 
absorbed on their way through the ejecta, giving rise to a flux deficit in the 
redshifted part of spectral emission lines. The required opacity might be generated 
by a multitude of weak Fe transitions. 

A surprisingly good agreement of the profiles of [O\,\textsc{i}] 
$\lambda\lambda6300,6364$ and Mg\,\textsc{i}] $\lambda4571$ (modified to account 
for the doublet nature of the oxygen feature) was found in most SNe regardless of 
their morphological class. This indicates that the line profiles are indeed 
determined by the ejecta geometry, and that Mg and O are similarly distributed 
within the SN ejecta. Deviations are mainly found in relatively early spectra, and 
we propose blending of the emerging nebular Mg\,\textsc{i}] emission with residual 
photospheric Fe\,\textsc{ii} lines as a possible reason for the differences.

\section*{Acknowledgments}

ST is grateful to Marilena Salvo for kindly agreeing to the use of a 
previously unpublished spectrum of SN~2002ap obtained at Siding Spring 
Observatory. He also wants to thank Bruno Leibundgut, Daniel Sauer, 
Luca Zampieri and Maryam Modjaz for stimulating discussions and helpful 
comments.
SB, EC and MT are supported by the Italian Ministry of Education 
via the PRIN 2006 n.2006022731 002 and ASI/INAF grant n. I/088/06/0. 
KM acknowledges the support by World Premier International Research 
Center Initiative, MEXT.

This work is based on data collected with the 3.6\,m, 2.2\,m, 1.5\,m 
and Danish Telescopes at ESO--La Silla (programme numbers 145.4-0004, 
057.D-0534, 058.D-0307, 059.D-0332, 060.D-0415, 061.D-0630, 065.H-0292 
and 066.D-0683), the 8.2\,m VLT-U1 at ESO--Paranal (programme numbers 
075.D-0662, 078.D-0246 and 079.D-0716), the 2.3\,m Telescope at Siding 
Spring Observatory and the 2.56\,m Nordic Optical Telescope at Roque de 
los Muchachos Observatory. 

The authors made use of the Asiago Supernova Catalogue, the NASA/IPAC 
Extragalactic Database (NED) which is operated by the Jet Propulsion 
Laboratory, California Institute of Technology, under contract with the 
National Aeronautics and Space Administration; the Lyon-Meudon Extragalactic 
Database (LEDA), supplied by the LEDA team at the Centre de Recherche 
Astronomique de Lyon, Observatoire de Lyon; the NIST Atomic Spectra 
Database, provided by the National Institute of Standards and Technology, 
Gaithersburg; the Online Supernova Spectrum Archive (SUSPECT), initiated 
and maintained at the Homer L. Dodge Department of Physics and Astronomy, 
University of Oklahoma; and the Bright Supernova web pages, maintained 
by David Bishop as part of the International Supernovae Network 
(http:/$\!$/www.supernovae.net).

\addcontentsline{toc}{chapter}{Bibliography}
\markboth{Bibliography}{Bibliography}
\bibliographystyle{mn2e}

\appendix

\section{Fit parameters}
\label{Fit parameters}

In the following tables the parameters of the best fits to the \OIa\ features 
of our spectra are reported. All numbers refer to the $\lambda6300$ line, since 
the $\lambda6364$ line is automatically added by the fitting algorithm at 1/3 
of the stength. One- and two-component fits have been obtained for all SNe. They 
are listed in Table~\ref{parameters12g}, along with error estimates for the 
one-component fit parameters. Three-component fits have only been performed for 
a few objets. In SNe~1990B, 2006ld and 2007C the line profiles were too complex 
to be satisfactorily reproduced with fewer components, while for SNe~1997dq, 
1998bw, 2002ap and 2006aj particular ejecta geometries were tested. The 
three-component fit parameters are reported in Table~\ref{parameters_3g}.

\begin{table*}
\caption{Parameters$^a$ of the best one-component (cols. 4--5) and two-component (cols. 6--12) 
         fits of [O\,\textsc{i}] $\lambda6300$ -- $1^{at}$ part.} 
\label{parameters12g}
\center
\begin{scriptsize}
\begin{tabular}{lclcccrccrcc}
\hline \\[-1.7ex]
\ \ SN  & Date & Sample region\quad\ & $\lambda$ & FWHM\quad\ \ & $\lambda_1$ & $\!\!\!$FWHM$_1\!$ & $\alpha_1$ & $\lambda_2$ & $\!\!\!$FWHM$_2\!$ & $\alpha_2$ & rel. RMS$^b$\\
\ \ (1) & (2)  & \qquad\ (3)         & (4)       & (5)\quad\ \  & (6)         & (7)\ \ \ \         & (8)        & (9)         & (10)\ \ \          & (11)       & (12)        \\[0.3ex] 
\hline \\[-1.7ex]
1983N  & 1983/03/01 & 6195.9--6455.8 & $6300.8\pm~2.7$ & $~97.1\pm~1.5$\quad\ \ &  6297.3  &  100.6\ \ \ &   0.91   &  6320.9  &   46.0\ \ \ &   0.09  &  0.545 \\
1985F  & 1985/03/19 & 6165.0--6477.0 & $6294.7\pm~1.7$ & $~87.1\pm~3.0$\quad\ \ &  6290.7  &  112.0\ \ \ &   0.76   &  6300.4  &   40.4\ \ \ &   0.24  &  0.062 \\
1987M  & 1988/02/09 & 6125.9--6530.6 & $6284.9\pm~3.1$ & $151.8\pm~3.0$\quad\ \ &  6283.3  &  158.8\ \ \ &   0.95   &  6300.4  &   31.6\ \ \ &   0.05  &  0.494 \\
       & 1988/02/25 & 6114.3--6495.3 & $6284.5\pm~4.4$ & $141.0\pm~9.0$\quad\ \ &  6282.8  &  149.4\ \ \ &   0.94   &  6297.6  &   37.6\ \ \ &   0.06  &  0.774 \\
1988L  & 1988/07/17 & 6114.2--6491.7 & $6266.3\pm~6.4$ & $157.7\pm13.0$\quad\ \ &    -     &    -  \ \ \ &    -     &    -     &    -  \ \ \ &    -    & -\ \ \ \\
       & 1988/09/15 & 6147.8--6467.0 & $6285.7\pm10.3$ & $147.1\pm30.0$\quad\ \ &    -     &    -  \ \ \ &    -     &    -     &    -  \ \ \ &    -    & -\ \ \ \\
1990B  & 1990/04/18 & 6132.1--6499.7 & $6285.5\pm~3.5$ & $183.2\pm~3.0$\quad\ \ &  6305.1  &  185.8\ \ \ &   0.84   &  6235.6  &   36.6\ \ \ &   0.16  &  0.230 \\
       & 1990/04/30 & 6192.4--6499.6 & $6292.5\pm~6.0$ & $194.5\pm12.0$\quad\ \ &  6299.4  &  198.7\ \ \ &   0.95   &  6242.3  &   16.7\ \ \ &   0.05  &  0.688 \\
1990I  & 1990/07/26 & 6114.5--6563.7 & $6287.3\pm~2.5$ & $184.4\pm~3.0$\quad\ \ &  6309.4  &  192.1\ \ \ &   0.76   &  6250.8  &   86.3\ \ \ &   0.24  &  0.345 \\
       & 1990/12/21 & 6171.2--6458.0 & $6300.2\pm~2.5$ & $106.4\pm~2.0$\quad\ \ &  6297.1  &  119.0\ \ \ &   0.89   &  6310.5  &   30.0\ \ \ &   0.11  &  0.454 \\
       & 1991/02/20 & 6178.6--6466.8 & $6298.5\pm~6.4$ & $113.5\pm15.0$\quad\ \ &  6296.1  &  125.4\ \ \ &   0.90   &  6307.1  &   24.9\ \ \ &   0.10  &  0.837 \\
1990U  & 1990/10/20 & 6189.8--6443.9 & $6283.5\pm~4.1$ & $~91.3\pm~6.0$\quad\ \ &  6267.3  &   63.7\ \ \ &   0.63   &  6319.2  &   73.1\ \ \ &   0.37  &  0.578 \\
       & 1990/10/24 & 6197.7--6435.5 & $6287.0\pm~4.1$ & $~88.2\pm~3.6$\quad\ \ &  6268.0  &   54.1\ \ \ &   0.57   &  6319.8  &   65.8\ \ \ &   0.43  &  0.349 \\
       & 1990/11/23 & 6183.1--6463.6 & $6292.6\pm~3.7$ & $~89.6\pm~2.0$\quad\ \ &  6270.9  &   49.6\ \ \ &   0.49   &  6319.8  &   68.8\ \ \ &   0.51  &  0.213 \\
       & 1990/11/28 & 6203.2--6457.3 & $6295.3\pm~5.3$ & $~89.7\pm~9.0$\quad\ \ &  6270.0  &   47.9\ \ \ &   0.45   &  6319.7  &   69.1\ \ \ &   0.55  &  0.373 \\
       & 1990/12/12 & 6195.7--6451.1 & $6291.5\pm~5.3$ & $~96.4\pm~6.0$\quad\ \ &  6268.6  &   54.3\ \ \ &   0.47   &  6317.6  &   77.5\ \ \ &   0.53  &  0.619 \\
       & 1990/12/20 & 6215.4--6448.5 & $6292.2\pm~3.8$ & $~92.8\pm~2.0$\quad\ \ &  6269.0  &   48.8\ \ \ &   0.50   &  6321.1  &   66.4\ \ \ &   0.50  &  0.255 \\
       & 1991/01/06 & 6193.9--6443.0 & $6294.6\pm~3.8$ & $~93.3\pm~5.0$\quad\ \ &  6270.5  &   48.9\ \ \ &   0.46   &  6322.3  &   71.6\ \ \ &   0.54  &  0.268 \\
       & 1991/01/12 & 6198.4--6465.8 & $6293.1\pm~3.9$ & $~93.9\pm~4.0$\quad\ \ &  6270.3  &   52.4\ \ \ &   0.50   &  6323.3  &   73.9\ \ \ &   0.50  &  0.309 \\
1990W  & 1991/02/21 & 6168.0--6449.5 & $6295.1\pm~3.0$ & $106.9\pm~3.0$\quad\ \ &  6295.6  &  120.2\ \ \ &   0.90   &  6292.7  &   40.6\ \ \ &   0.10  &  0.499 \\
       & 1991/04/21 & 6144.0--6496.8 & $6295.5\pm~3.0$ & $107.9\pm~1.0$\quad\ \ &  6298.0  &  125.1\ \ \ &   0.84   &  6289.7  &   46.5\ \ \ &   0.16  &  0.290 \\
1990aa & 1991/01/12 & 6128.2--6516.7 & $6283.3\pm~3.8$ & $184.4\pm~8.8$\quad\ \ &  6306.8  &  191.0\ \ \ &   0.79   &  6236.5  &   77.5\ \ \ &   0.21  &  0.618 \\
       & 1991/01/23 & 6128.5--6504.9 & $6283.3\pm~4.3$ & $165.8\pm13.0$\quad\ \ &  6300.5  &  164.7\ \ \ &   0.79   &  6238.4  &   96.2\ \ \ &   0.21  &  0.912 \\
1990aj & 1991/01/29 & 6150.1--6498.6 & $6294.1\pm~2.0$ & $142.4\pm~3.0$\quad\ \ &  6292.2  &  162.7\ \ \ &   0.88   &  6300.6  &   46.2\ \ \ &   0.12  &  0.491 \\
       & 1991/02/22 & 6170.4--6450.2 & $6292.9\pm~2.0$ & $115.4\pm~5.0$\quad\ \ &  6289.7  &  141.1\ \ \ &   0.83   &  6299.8  &   40.6\ \ \ &   0.17  &  0.287 \\
       & 1991/03/10 & 6149.3--6492.6 & $6296.6\pm~8.2$ & $124.2\pm19.0$\quad\ \ &  6290.8  &  158.6\ \ \ &   0.77   &  6304.6  &   44.3\ \ \ &   0.23  &  0.706 \\
1991A  & 1991/03/22 & 6134.9--6467.6 & $6264.1\pm~3.5$ & $112.5\pm~5.5$\quad\ \ &  6266.0  &  116.2\ \ \ &   0.95   &  6246.6  &   34.5\ \ \ &   0.05  &  0.671 \\
       & 1991/04/07 & 6137.7--6469.1 & $6276.1\pm~5.4$ & $124.4\pm11.0$\quad\ \ &  6276.8  &  127.7\ \ \ &   0.98   &  6262.0  &   18.9\ \ \ &   0.02  &  0.899 \\
       & 1991/04/16 & 6135.6--6480.0 & $6276.4\pm~2.2$ & $120.8\pm~3.0$\quad\ \ &  6278.5  &  125.3\ \ \ &   0.94   &  6258.1  &   39.8\ \ \ &   0.06  &  0.647 \\
       & 1991/06/08 & 6137.1--6508.2 & $6286.5\pm~2.0$ & $126.1\pm~3.0$\quad\ \ &  6289.1  &  130.4\ \ \ &   0.95   &  6263.5  &   31.7\ \ \ &   0.05  &  0.340 \\
1991L  & 1991/06/08 & 6156.3--6444.7 & $6289.5\pm~5.5$ & $105.2\pm~6.5$\quad\ \ &  6258.9  &   65.5\ \ \ &   0.50   &  6320.4  &   67.6\ \ \ &   0.50  &  0.694 \\
1991N  & 1991/12/14 & 6191.9--6457.6 & $6303.9\pm~3.6$ & $106.1\pm~8.0$\quad\ \ &  6303.7  &  106.1\ \ \ &   0.77   &  6304.3  &  105.9\ \ \ &   0.23  &  1.048 \\
       & 1992/01/09 & 6191.0--6472.7 & $6303.5\pm~2.8$ & $104.0\pm~5.0$\quad\ \ &  6304.5  &  123.9\ \ \ &   0.68   &  6302.8  &   74.9\ \ \ &   0.32  &  0.943 \\
1993J$^c$ &1993/10/19 &6172.2--6512.2 &$6279.7\pm~4.2$ & $110.0\pm~2.0$\quad\ \ &  6307.1  &  115.7\ \ \ &   0.55   &  6257.6  &   73.4\ \ \ &   0.45  &  0.525 \\
       & 1993/11/19 & 6192.2--6512.2 & $6294.3\pm~4.3$ & $110.0\pm~2.0$\quad\ \ &  6307.3  &  114.4\ \ \ &   0.77   &  6265.3  &   54.2\ \ \ &   0.23  &  0.404 \\
       & 1993/12/08 & 6195.7--6512.2 & $6295.8\pm~4.2$ & $110.1\pm~2.0$\quad\ \ &  6308.4  &  109.2\ \ \ &   0.79   &  6264.7  &   54.4\ \ \ &   0.21  &  0.460 \\
       & 1994/01/17 & 6176.4--6512.0 & $6292.6\pm~4.2$ & $117.9\pm~2.0$\quad\ \ &  6304.2  &  110.1\ \ \ &   0.79   &  6261.8  &   54.2\ \ \ &   0.21  &  0.500 \\
       & 1994/01/21 & 6188.6--6513.1 & $6289.2\pm~4.2$ & $120.0\pm~1.5$\quad\ \ &  6303.8  &  114.7\ \ \ &   0.72   &  6261.9  &   63.4\ \ \ &   0.28  &  0.421 \\
       & 1994/01/22 & 6178.6--6514.3 & $6293.4\pm~4.7$ & $111.0\pm~4.0$\quad\ \ &  6305.2  &  110.0\ \ \ &   0.78   &  6262.1  &   54.3\ \ \ &   0.22  &  0.469 \\
       & 1994/03/25 & 6183.9--6514.1 & $6295.9\pm~4.6$ & $115.6\pm~4.0$\quad\ \ &  6306.4  &  108.9\ \ \ &   0.80   &  6265.7  &   57.1\ \ \ &   0.20  &  0.574 \\
       & 1994/03/30 & 6191.3--6518.9 & $6302.3\pm~4.6$ & $110.0\pm~6.0$\quad\ \ &  6309.7  &  104.0\ \ \ &   0.84   &  6269.5  &   50.5\ \ \ &   0.16  &  0.635 \\
1994I  & 1994/07/14 & 6130.9--6497.4 & $6275.6\pm~2.0$ & $140.3\pm~2.0$\quad\ \ &  6294.6  &  129.2\ \ \ &   0.77   &  6231.6  &   62.4\ \ \ &   0.23  &  0.169 \\
       & 1994/08/04 & 6140.0--6493.2 & $6279.9\pm~2.7$ & $138.3\pm~2.0$\quad\ \ &  6296.3  &  127.5\ \ \ &   0.80   &  6231.2  &   62.0\ \ \ &   0.20  &  0.209 \\
       & 1994/09/02 & 6160.3--6469.5 & $6277.4\pm~2.7$ & $129.3\pm~4.0$\quad\ \ &  6294.9  &  111.3\ \ \ &   0.77   &  6229.7  &   60.0\ \ \ &   0.23  &  0.289 \\
1995bb & 1995/12/17 & 6157.7--6470.4 & $6296.6\pm~6.9$ & $145.0\pm12.0$\quad\ \ &  6292.3  &  152.3\ \ \ &   0.93   &  6321.8  &   45.6\ \ \ &   0.07  &  0.929 \\
1996D  & 1996/09/10 & 6150.8--6483.0 & $6301.3\pm~7.9$ & $152.2\pm15.0$\quad\ \ &  6288.6  &  158.0\ \ \ &   0.87   &  6337.4  &   39.6\ \ \ &   0.13  &  0.767 \\
1996N  & 1996/10/19 & 6164.7--6471.2 & $6283.7\pm~5.0$ & $114.1\pm~5.0$\quad\ \ &  6299.5  &   94.4\ \ \ &   0.77   &  6239.0  &   46.1\ \ \ &   0.23  &  0.414 \\
       & 1996/12/16 & 6177.7--6465.1 & $6281.1\pm~5.0$ & $107.6\pm~5.0$\quad\ \ &  6295.6  &   89.6\ \ \ &   0.78   &  6237.4  &   47.0\ \ \ &   0.22  &  0.449 \\
       & 1997/01/13 & 6174.5--6458.2 & $6282.2\pm~5.0$ & $109.5\pm~5.0$\quad\ \ &  6294.9  &   94.9\ \ \ &   0.80   &  6236.7  &   51.2\ \ \ &   0.20  &  0.754 \\
       & 1997/02/12 & 6173.0--6453.2 & $6282.6\pm~5.0$ & $103.4\pm~5.0$\quad\ \ &  6300.4  &   81.8\ \ \ &   0.72   &  6245.1  &   45.8\ \ \ &   0.28  &  0.389 \\
1996aq & 1997/02/11 & 6140.1--6494.1 & $6286.6\pm~2.2$ & $128.2\pm~2.0$\quad\ \ &  6300.7  &  112.5\ \ \ &   0.83   &  6238.5  &   25.4\ \ \ &   0.17  &  0.104 \\
       & 1997/04/02 & 6165.2--6468.4 & $6288.5\pm~3.6$ & $123.7\pm~6.0$\quad\ \ &  6299.5  &  112.2\ \ \ &   0.86   &  6241.8  &   19.9\ \ \ &   0.14  &  0.090 \\
       & 1997/05/14 & 6143.0--6508.3 & $6291.5\pm~2.2$ & $126.0\pm~1.0$\quad\ \ &  6302.1  &  114.2\ \ \ &   0.87   &  6240.8  &   19.6\ \ \ &   0.13  &  0.073 \\
1997B  & 1997/09/23 & 6186.0--6438.9 & $6301.0\pm~8.6$ & $110.7\pm10.0$\quad\ \ &  6295.5  &   93.6\ \ \ &   0.63   &  6317.7  &  143.0\ \ \ &   0.37  &  1.011 \\
       & 1997/10/11 & 6206.4--6432.3 & $6296.7\pm~7.7$ & $~96.2\pm15.0$\quad\ \ &  6280.6  &   83.1\ \ \ &   0.56   &  6318.8  &   86.0\ \ \ &   0.44  &  1.048 \\
       & 1998/02/02 & 6201.3--6481.6 & $6304.6\pm~5.9$ & $~97.3\pm15.0$\quad\ \ &  6286.3  &   71.1\ \ \ &   0.37   &  6318.4  &   95.2\ \ \ &   0.63  &  0.968 \\
1997X  & 1997/05/10 & 6156.4--6495.1 & $6295.8\pm~2.2$ & $106.3\pm~4.0$\quad\ \ &  6298.4  &  146.5\ \ \ &   0.53   &  6294.7  &   77.0\ \ \ &   0.47  &  0.693 \\
       & 1997/05/13 & 6143.4--6517.7 & $6297.2\pm~1.5$ & $122.6\pm~3.0$\quad\ \ &  6295.3  &  145.3\ \ \ &   0.76   &  6300.2  &   72.2\ \ \ &   0.24  &  0.566 \\
1997dq & 1998/05/30 & 6205.0--6446.5 & $6300.0\pm~2.8$ & $~97.8\pm~1.0$\quad\ \ &  6323.7  &   68.7\ \ \ &   0.55   &  6271.2  &   63.6\ \ \ &   0.45  &  0.592 \\
       & 1998/06/18 & 6202.2--6439.8 & $6298.8\pm~4.0$ & $~98.6\pm~7.0$\quad\ \ &  6320.5  &   71.5\ \ \ &   0.55   &  6270.2  &   72.7\ \ \ &   0.45  &  0.840 \\
1997ef & 1998/09/21 & 6166.1--6470.1 & $6294.9\pm~5.3$ & $120.2\pm15.0$\quad\ \ &  6300.6  &  179.2\ \ \ &   0.72   &  6290.8  &   44.8\ \ \ &   0.28  &  0.486 \\
1998bw & 1998/09/12 & 6090.0--6549.7 & $6298.4\pm~5.8$ & $185.5\pm~6.0$\quad\ \ &  6297.3  &  191.2\ \ \ &   0.97   &  6315.1  &   31.6\ \ \ &   0.03  &  0.730 \\
       & 1998/11/26 & 6135.6--6504.6 & $6298.4\pm~3.2$ & $146.1\pm~4.0$\quad\ \ &  6292.8  &  166.1\ \ \ &   0.87   &  6315.1  &   42.6\ \ \ &   0.13  &  0.124 \\
       & 1999/04/12 & 6134.6--6500.3 & $6299.3\pm~3.4$ & $120.6\pm~5.0$\quad\ \ &  6294.0  &  142.8\ \ \ &   0.84   &  6310.8  &   39.1\ \ \ &   0.16  &  0.107 \\
       & 1999/05/21 & 6162.3--6460.7 & $6297.6\pm~3.2$ & $109.8\pm~3.0$\quad\ \ &  6291.7  &  132.7\ \ \ &   0.82   &  6308.2  &   39.0\ \ \ &   0.18  &  0.163 \\
1999cn & 2000/08/04 & 6198.2--6421.9 & $6294.6\pm10.3$ & $108.4\pm20.0$\quad\ \ &  6292.0  &  123.0\ \ \ &   0.96   &  6303.7  &   32.9\ \ \ &   0.04  &  0.785 \\
1999dn & 2000/09/01 & 6214.1--6455.2 & $6309.6\pm10.2$ & $~94.3\pm20.0$\quad\ \ &  6304.1  &  112.9\ \ \ &   0.81   &  6322.9  &   33.8\ \ \ &   0.19  &  0.920 \\[0.3ex]
\hline
\end{tabular}
\end{scriptsize}
\end{table*}

\addtocounter{table}{-1}
\begin{table*}
\caption[]{\textit{cont.} Parameters$^a$ of the best one-component (cols. 4--5) and two-component (cols. 6--12) fits of [O\,\textsc{i}] $\lambda6300$ -- $2^{nd}$ part.}
\label{}
\center
\begin{scriptsize}
\begin{tabular}{lclcccrccrcc}
\hline \\[-1.7ex]
\ \ SN  & Date & Sample region\quad\ & $\lambda$ & FWHM\quad\ \ & $\lambda_1$ & $\!\!\!$FWHM$_1\!$ & $\alpha_1$ & $\lambda_2$ & $\!\!\!$FWHM$_2\!$ & $\alpha_2$ & rel. RMS$^b$\\
\ \ (1) & (2)  & \qquad\ (3)         & (4)       & (5)\quad\ \  & (6)         & (7)\ \ \ \         & (8)        & (9)         & (10)\ \ \          & (11)       & (12)        \\[0.3ex] 
\hline \\[-1.7ex]
2000ew & 2001/03/17 & 6189.4--6420.2 & $6284.3\pm~3.0$ & $~99.3\pm~5.0$\quad\ \ &  6294.3  &  109.6\ \ \ &   0.79   &  6266.3  &   33.0\ \ \ &   0.21  &  0.496 \\
2002ap & 2002/08/06 & 6082.7--6529.4 & $6297.9\pm~2.3$ & $172.5\pm~5.0$\quad\ \ &  6296.6  &  179.2\ \ \ &   0.96   &  6312.3  &   21.9\ \ \ &   0.04  &  0.615 \\
       & 2002/06/17 & 6064.6--6516.5 & $6300.0\pm~1.5$ & $174.0\pm~3.0$\quad\ \ &  6298.7  &  181.5\ \ \ &   0.95   &  6312.5  &   22.5\ \ \ &   0.05  &  0.254 \\
       & 2002/07/11 & 6084.1--6514.9 & $6302.4\pm~1.5$ & $163.1\pm~1.0$\quad\ \ &  6301.6  &  164.6\ \ \ &   0.97   &  6313.0  &   20.0\ \ \ &   0.03  &  0.324 \\
       & 2002/08/09 & 6103.9--6503.4 & $6304.6\pm~1.5$ & $157.5\pm~1.0$\quad\ \ &  6303.5  &  165.7\ \ \ &   0.96   &  6314.1  &   21.6\ \ \ &   0.04  &  0.164 \\
       & 2002/10/01 & 6094.6--6513.3 & $6305.4\pm~1.5$ & $150.4\pm~2.0$\quad\ \ &  6304.1  &  157.9\ \ \ &   0.95   &  6314.4  &   20.2\ \ \ &   0.05  &  0.216 \\
       & 2002/10/09 & 6104.7--6511.7 & $6305.0\pm~1.5$ & $152.2\pm~4.0$\quad\ \ &  6303.9  &  160.6\ \ \ &   0.95   &  6313.5  &   20.3\ \ \ &   0.05  &  0.366 \\
       & 2002/10/14 & 6104.0--6511.2 & $6306.4\pm~1.5$ & $153.9\pm~3.0$\quad\ \ &  6304.2  &  166.4\ \ \ &   0.93   &  6316.2  &   31.0\ \ \ &   0.07  &  0.435 \\
       & 2002/11/06 & 6149.5--6500.5 & $6305.7\pm~1.5$ & $149.0\pm~1.0$\quad\ \ &  6304.7  &  157.8\ \ \ &   0.95   &  6312.5  &   22.2\ \ \ &   0.05  &  0.247 \\
       & 2003/01/07 & 6137.7--6519.0 & $6306.9\pm~2.3$ & $143.0\pm~2.0$\quad\ \ &  6305.9  &  150.8\ \ \ &   0.95   &  6313.7  &   21.7\ \ \ &   0.05  &  0.373 \\
       & 2003/02/27 & 6145.0--6490.0 & $6309.5\pm~1.5$ & $140.0\pm~2.0$\quad\ \ &  6309.0  &  140.8\ \ \ &   0.97   &  6315.4  &   17.1\ \ \ &   0.03  &  0.534 \\
2003jd & 2004/09/11 & 6087.3--6505.8 & $6299.1\pm~3.6$ & $219.6\pm~8.0$\quad\ \ &  6232.0  &  116.7\ \ \ &   0.50   &  6357.2  &   87.0\ \ \ &   0.50  &  0.435 \\
       & 2004/10/18 & 6150.7--6488.2 & $6304.4\pm~6.3$ & $187.5\pm15.0$\quad\ \ &  6251.4  &   87.0\ \ \ &   0.52   &  6358.5  &   67.8\ \ \ &   0.48  &  0.584 \\
2004aw & 2004/11/14 & 6124.1--6498.6 & $6296.1\pm~3.3$ & $118.7\pm~7.0$\quad\ \ &  6293.1  &  133.0\ \ \ &   0.89   &  6307.6  &   37.0\ \ \ &   0.11  &  0.507 \\
       & 2004/12/08 & 6191.6--6430.1 & $6289.2\pm~6.6$ & $~91.0\pm14.0$\quad\ \ &  6287.5  &   94.7\ \ \ &   0.94   &  6302.9  &   34.8\ \ \ &   0.06  &  0.882 \\
       & 2005/05/11 & 6199.0--6442.0 & $6298.0\pm~6.6$ & $~91.4\pm16.0$\quad\ \ &  6294.0  &  101.8\ \ \ &   0.87   &  6310.8  &   32.7\ \ \ &   0.13  &  0.651 \\
2004gt & 2005/05/24 & 6174.8--6485.3 & $6285.2\pm~3.1$ & $125.1\pm~3.0$\quad\ \ &  6260.7  &   72.5\ \ \ &   0.60   &  6335.8  &   87.1\ \ \ &   0.40  &  0.490 \\
2005N  & 2005/01/21 & 6160.0--6476.3 & $6268.6\pm~6.5$ & $117.3\pm~7.5$\quad\ \ &  6284.0  &  126.3\ \ \ &   0.77   &  6241.2  &   46.8\ \ \ &   0.23  &  0.353 \\
2006F  & 2006/11/16 & 6223.0--6450.9 & $6296.9\pm~6.3$ & $~75.8\pm20.0$\quad\ \ &  6314.6  &   84.7\ \ \ &   0.64   &  6281.4  &   38.9\ \ \ &   0.36  &  0.508 \\
2006T  & 2007/02/18 & 6203.0--6452.4 & $6300.5\pm~8.0$ & $123.6\pm16.0$\quad\ \ &  6266.8  &   50.7\ \ \ &   0.51   &  6338.0  &   57.9\ \ \ &   0.49  &  0.306 \\
2006aj & 2006/09/19 & 6101.0--6498.7 & $6294.6\pm~6.1$ & $188.5\pm~9.0$\quad\ \ &  6295.0  &  192.2\ \ \ &   0.99   &  6301.8  &   18.1\ \ \ &   0.01  &  0.891 \\
       & 2006/11/27 & 6096.0--6515.7 & $6303.5\pm~6.7$ & $203.4\pm11.0$\quad\ \ &  6304.5  &  222.4\ \ \ &   0.93   &  6300.8  &   35.3\ \ \ &   0.07  &  0.588 \\
       & 2006/12/19 & 6119.2--6507.0 & $6305.2\pm~7.7$ & $177.1\pm12.0$\quad\ \ &  6306.2  &  197.4\ \ \ &   0.92   &  6302.3  &   22.4\ \ \ &   0.08  &  0.410 \\
2006gi & 2007/02/10 & 6138.3--6433.0 & $6278.5\pm~4.0$ & $107.0\pm~2.0$\quad\ \ &  6280.1  &  114.5\ \ \ &   0.93   &  6267.7  &   35.2\ \ \ &   0.07  &  0.458 \\
2006ld & 2007/07/17 & 6176.4--6492.5 & $6285.5\pm~2.2$ & $~97.2\pm~5.5$\quad\ \ &  6304.8  &  104.0\ \ \ &   0.70   &  6259.7  &   43.9\ \ \ &   0.30  &  0.572 \\
       & 2007/08/06 & 6173.5--6491.2 & $6287.6\pm~3.1$ & $102.1\pm~9.0$\quad\ \ &  6305.7  &  115.1\ \ \ &   0.72   &  6263.5  &   43.8\ \ \ &   0.28  &  0.531 \\
       & 2007/08/20 & 6175.5--6482.4 & $6290.3\pm~2.6$ & $~92.7\pm~7.0$\quad\ \ &  6296.3  &   87.3\ \ \ &   0.89   &  6251.9  &   33.9\ \ \ &   0.11  &  0.936 \\
2007C  & 2007/05/17 & 6131.0--6501.4 & $6278.2\pm~3.6$ & $122.5\pm~2.0$\quad\ \ &  6301.1  &   92.3\ \ \ &   0.69   &  6235.0  &   52.4\ \ \ &   0.31  &  0.446 \\
       & 2007/06/20 & 6134.0--6502.5 & $6285.3\pm~2.6$ & $121.5\pm~1.5$\quad\ \ &  6299.6  &  103.2\ \ \ &   0.81   &  6237.6  &   41.6\ \ \ &   0.19  &  0.344 \\
2007I  & 2007/06/18 & 6157.1--6494.7 & $6300.3\pm~5.3$ & $118.8\pm~1.0$\quad\ \ &  6305.3  &  126.1\ \ \ &   0.90   &  6277.8  &   35.8\ \ \ &   0.10  &  0.122 \\
       & 2007/07/15 & 6159.2--6504.8 & $6303.4\pm~5.3$ & $111.9\pm~1.5$\quad\ \ &  6308.2  &  116.3\ \ \ &   0.90   &  6278.8  &   33.7\ \ \ &   0.10  &  0.129 \\[0.3ex]
\hline 
\end{tabular}
\flushleft
$^a$ Central wavelengths ($\lambda_i$), full widths at half maximum 
(FWHM$_i$) and relative strengths ($\alpha_i$) of the fit components 
(cf. Section~\ref{O_i_6300 and O_i_6364}). The sample region, $\lambda$ 
and FWHM are given in units of \AA.\\
$^b$ RMS of the fit, relative to the RMS of the one-component fit.\\
$^c$ The fitting was performed in H$\alpha$-subtracted spectra. For this purpose, 
a smoothed symmetric H$\alpha$ profile was subtracted, constructed by reflecting 
the red wing at the rest wavelength (see also \citealt{Patat95}).\\
\vspace*{0.6cm}
\end{scriptsize}
\end{table*}

\begin{table*}
\caption{Parameters$^a$ of the best three-component fits of [O\,\textsc{i}] $\lambda6300$ for 
        selected spectra with complex line profiles or particular geometric configurations.}
\label{parameters_3g}
\center
\begin{scriptsize}
\begin{tabular}{lclcclcclcclcl}
\hline \\[-1.7ex]
\ \ SN     & Date       & Sample region\quad\  & $\lambda_1$ & $\!\!$FWHM$_1\!\!$ & \ $\alpha_1$ & $\lambda_2$ & $\!\!$FWHM$_2\!\!$ & \ $\alpha_2$ & $\lambda_3$ & $\!\!$FWHM$_3\!\!$  & \ $\alpha_3$ & RMS$^b$ & Config.\\
\ \ (1)    &  (2)       &   \qquad\ (3)        &   (4)       &   (5)              & $\!$ (6)     &   (7)       &   (8)              & $\!$ (9)     &   (10)      &   (11)                         & (12)         &  (13)   & \ (14) \\[0.0cm] 
\hline \\[-1.7ex]
1990B  & 1990/04/18 & 6132.1--6499.7  &  6319.9  &   180.7  &    0.71\ \ &  6235.5  &   50.0   &    0.26\ \ &  6303.1  &   13.0   &    0.03\ \ &   0.140 &  clumpy    \\
1997dq & 1998/05/30 & 6205.0--6446.5  &  6272.4  &    63.3  &    0.49\ \ &  6326.9  &   63.3   &    0.49\ \ &  6298.4  &   12.3   &    0.02\ \ &   0.480 &  DP + NC   \\
1998bw & 1999/05/21 & 6162.3--6460.7  &  6259.4  &   118.2  &    0.38\ \ &  6319.4  &  118.2   &    0.38\ \ &  6307.8  &   45.0   &    0.24\ \ &   0.175 &  DP + NC   \\
2002ap & 2002/10/01 & 6094.6--6513.3  &  6251.8  &   102.6  &    0.42\ \ &  6352.1  &  102.6   &    0.42\ \ &  6313.5  &   42.3   &    0.15\ \ &   0.170 &  DP + NC   \\
2004gt & 2005/05/24 & 6174.8--6485.3  &  6243.8  &    63.7  &    0.37\ \ &  6338.6  &   63.7   &    0.37\ \ &  6279.3  &   44.3   &    0.26\ \ &   0.229 &  DP + blob   \\
2006aj & 2006/12/19 & 6119.2--6507.0  &  6256.9  &   151.4  &    0.45\ \ &  6353.3  &  151.4   &    0.45\ \ &  6302.2  &   24.1   &    0.09\ \ &   0.402 &  DP + NC   \\
2006ld & 2007/07/17 & 6176.4--6492.5  &  6320.8  &   121.4  &    0.48\ \ &  6261.6  &   40.1   &    0.40\ \ &  6301.6  &   20.0   &    0.12\ \ &   0.161 &  clumpy    \\
       & 2007/08/06 & 6173.5--6491.2  &  6320.4  &   136.6  &    0.53\ \ &  6264.2  &   39.4   &    0.36\ \ &  6302.1  &   21.1   &    0.11\ \ &   0.212 &  clumpy    \\
       & 2007/08/20 & 6175.5--6482.4  &  6322.3  &   117.5  &    0.47\ \ &  6264.4  &   46.5   &    0.37\ \ &  6303.2  &   19.9   &    0.16\ \ &   0.343 &  clumpy    \\ 
2007C  & 2007/05/17 & 6131.0--6501.4  &  6299.0  &   127.1  &    0.56\ \ &  6239.5  &   54.2   &    0.29\ \ &  6301.1  &   43.1   &    0.15\ \ &   0.088 &  NC + blob \\
       & 2007/06/20 & 6134.0--6502.5  &  6297.7  &   125.6  &    0.68\ \ &  6242.0  &   43.5   &    0.19\ \ &  6304.4  &   41.4   &    0.13\ \ &   0.060 &  NC + blob \\[0.0cm]
\hline
\end{tabular}
\flushleft
$^a$ Central wavelengths ($\lambda_i$), full widths at half maximum 
(FWHM$_i$) and relative strengths ($\alpha_i$) of the fit components 
(cf. Section~\ref{O_i_6300 and O_i_6364}). The sample region, $\lambda$ 
and FWHM are given in units of \AA.\\
$^b$ RMS of the fit, relative to the RMS of the one-component fit (cf. Table~\ref{parameters12g}).\\
\end{scriptsize}
\end{table*}


\label{lastpage}

\end{document}